\documentclass{article}
\usepackage{amssymb}
\usepackage{amsfonts}
\usepackage{graphicx}
\usepackage{amsmath}

\newtheorem{theorem}{Theorem}

\newtheorem{corollary}[theorem]{Corollary}

\newtheorem{lemma}[theorem]{Lemma}

\newtheorem{proposition}[theorem]{Proposition}

\makeatletter \@addtoreset{equation}{section} \makeatother
\numberwithin{equation}{section} \numberwithin{theorem}{section}

\def\ds{\displaystyle}

\def\fracd{\displaystyle\frac}

\begin{document}

\title{On the edge universality of the local eigenvalue statistics of matrix
models}
\author{L. Pastur$^{1,2}$, and M. Shcherbina$^{2}$ \\
$^1$ University Paris 7, Paris, France\\
$^2$ Institute for Low Temperature Physics, Kharkov, Ukraine}
\maketitle
\date{}

\begin{abstract}
Basing on our recent results on the $1/n$-expansion in unitary invariant
random matrix ensembles, known as matrix models, we prove that the local
eigenvalue statistic, arising in a certain neighborhood of the edges of the
support of the Density of States, is independent of the form of the
potential, determining the matrix model. Our proof is applicable to the case
of real analytic potentials and of supports, consisting of one or two
disjoint intervals.
\end{abstract}

\bigskip

\section{Introduction}

Universality is an important concept of the random matrix theory and of its
numerous applications (see e.g. reviews \cite{We-Co:98,Pa:00} and references
therein). In a more concrete context one refers to universality while
dealing \ with local eigenvalue statistics of ensembles of $n\times n$ real
symmetric, Hermitian or real quaternion matrices in the limit $n\rightarrow
\infty $. One distinguishes the bulk case, arising in a $1/n$-neighborhood
of a point $\lambda _{0}$ of the support of the Density of States $\rho $ of
an ensemble, such that $0<\rho (\lambda _{0})<\infty $, and the edge case,
arising in a certain $o(1)$-neighborhood of endpoints of the support, more
generally, in a neighborhood of those points of the support, at which $\rho
(\lambda _{0})=0,\infty $ (perhaps as an one-side limit, i.e., for $\lambda
_{0}+0$ or $\lambda _{0}-0$).

In this paper we will study the edge universality of ensembles of Hermitian
matrices $M=\{M_{jk}\}_{j,k=1}^n$, $\overline M_{jk}=M_{j,k}$, known as the
matrix models and defined by the probability distribution

\begin{equation}
P_{n}(M)dM=Z_{n}^{-1}\exp \{-n\hbox{Tr}V(M)\}dM,  \label{p(M)}
\end{equation}%
where
\begin{equation}
dM=\prod_{j=1}^{n}dM_{jj}\prod_{1\leq j<k\leq n}d\Im M_{jk}d\Re
M_{jk}, \label{dM}
\end{equation}%
$Z_{n}$ is the normalizing constant, and the function $V:\mathbf{R}%
\rightarrow \mathbf{R}_{+}$ is called the potential. We assume that $V$
satisfies the conditions:

\begin{enumerate}
\item[(i)] there exist $L_{1}$ and $\epsilon >0$, such that
\begin{equation}
|V(\lambda )|\geq (2+\epsilon )\log |\lambda |,\,|\lambda |\geq L_{1},
\label{log}
\end{equation}

\item[(ii)] for any $0<L_{2}<\infty $ and some $\gamma >0$
\begin{equation}
|V(\lambda _{1})-V(\lambda _{2})|\leq C(L_{2})|\lambda _{1}-\lambda
_{2}|^{\gamma },\ \ |\lambda _{1,2}|\leq L_{2}.  \label{Lip}
\end{equation}
\end{enumerate}
Denote by $\lambda _{1}^{(n)},...,\lambda _{n}^{(n)}$ the eigenvalues of a
matrix $M_{n}$ and define its Eigenvalue Counting Measure as
\begin{equation}
N_{n}(\Delta )=\sharp \{\lambda _{l}^{(n)}\in \Delta ,\,\,\,l=1,\dots
,n\}\cdot n^{-1},  \label{NCM}
\end{equation}%
where $\Delta $ is an interval of the spectral axis. According to \cite%
{BPS,Jo:98} the $N_n$ tends weakly in probability as $n\rightarrow \infty $ to
the non-random measure $N$ known as the Integrated Density of States (IDS)
of the ensemble. The measure $N$ is a unique minimizer of the functional
\begin{equation}
\mathcal{E}[m]=\int V(\lambda )m(d\lambda )-\int \int \log |\lambda -\mu
|m(d\lambda )m(d\mu ),  \label{En}
\end{equation}%
defined on non-negative unit measures on $\mathbf{R}$. Here and below
integrals without limits denote the integration over the whole real axis.

The IDS $N$ is normalized to unity and is absolutely continuous if $%
V^{\prime }$ satisfies the Lipshitz condition \cite{Sa-To:97}:
\begin{equation}
N(\mathbf{R})=1,\;N(\Delta )=\int_{\Delta }\rho (\lambda )d\lambda .
\label{Nrho}
\end{equation}%
The non-negative function $\rho $ in (\ref{Nrho}) is called the Density of
States (DOS) of the ensemble. The DOS of matrix models was studied in \cite%
{BPS,Jo:98,De-Co:98}. It follows from these papers that in the case of a
real analytic potential the support of the DOS consists of a finite number
of finite disjoint intervals and that if $a_{\ast }$ is an endpoint of the
support, then the DOS behaves asymptotically as $\rho (\lambda )=\mathrm{%
const}\cdot |\lambda -a_{\ast }|^{1/2},\;\lambda \rightarrow a_{\ast }$
generically in $V$.

The most studied ensemble of the random matrix theory is the Gaussian
Unitary Ensemble, determined by (\ref{p(M)}) -\ (\ref{dM}) with
\begin{equation}
V(\lambda )=2\lambda ^{2}/a^{2}.  \label{Ga}
\end{equation}%
Here the DOS is the semi-circle low of Wigner
\begin{equation}
\rho (\lambda )=\frac{2}{\pi a^{2}}(a^{2}-\lambda ^{2})_{+}^{1/2},
\label{scl}
\end{equation}%
where $x_{+}=\max (x,0)$.

The most known quantity probing the universality is the large-$n$ form of
the hole probability
\begin{equation}
E_{n}(\Delta _{n})=\mathbf{P}_{n}\{\lambda _{l}^{(n)}\notin \Delta
_{n},\;l=1,...,n\},  \label{hon}
\end{equation}%
where $\mathbf{P}_{n}\{...\}$ is the probability defined by the distribution
(\ref{p(M)}) -\ (\ref{dM}), and $\Delta _{n}$ is an interval of the spectral
axis, whose order of magnitude is fixed by the condition $nN(\Delta
_{n})|\Delta _{n}|\eqsim 1$.

In the bulk case we choose \cite{Me:91}
\begin{equation}
\Delta _{n}=(\lambda _{0},\lambda _{0}+s/n\rho (\lambda
_{0})),\;s>0,\;0<\rho (\lambda _{0})<\infty .  \label{Deb}
\end{equation}
In this case the limiting hole probability is the Fredholm determinant of
the integral operator, defined by the kernel $\sin \pi (t_{1}-t_{2})/\pi
(t_{1}-t_{2})$ on the interval $(0,s)$. This fact for the Gaussian Unitary
Ensemble (\ref{Ga}) was established by M. Gaudin in the early 60s \cite%
{Me:91}. The same fact was proved recently in \cite{PS:97,De-Co:99} for
certain classes of matrix models. This is an example of bulk universality,
showing that the local (in the sense (\ref{hon}) - (\ref{Deb})) statistical
properties of eigenvalues do not depend on the ensemble, i.e., on the
function $V$ in (\ref{p(M)}), modulo a proper rescaling of the spectral axis.

The edge case of local eigenvalue statistics was studied much later even for
the GUE \cite{Fo:93,Tr-Wi:94a}. It was found that if we choose
\begin{equation}
\Delta _{n}=\left( a,a(1+s/2n^{2/3})\right) ,\;s\in \mathbf{R}  \label{DeA}
\end{equation}%
for the right-hand edge of the support $[-a,a]$ of (\ref{scl}), then the
limit as $n\rightarrow \infty $ of the hole probability (\ref{hon}) of the
GUE is the Fredholm determinant of the integral operator, defined on the
interval $(0,s)$ by the kernel
\begin{equation}
\mathcal{K}(t_{1},t_{2})=\frac{Ai(t_{1})Ai^{\prime }(t_{2})-Ai^{\prime
}(t_{1})Ai(t_{2})}{t_{1}-t_{2}},  \label{Aik}
\end{equation}%
where $Ai$ is the standard Airy function \cite{AS}. Similar result is valid
for the left-hand edge of the support of (\ref{scl}). Hence, the edge
universality means that if $a_{\ast }$ is an endpoint of the DOS\ support,
and $\rho $ behaves asymptotically as $\rho (\lambda )=\mathrm{const}\cdot
|\lambda -a_{\ast }|^{1/2},\;\lambda \rightarrow a_{\ast }$, then the
limiting hole probability should be the same Fredholm determinant.

This fact for real analytic potentials in (\ref{p(M)}) can be deduced, under
certain conditions, from the recent results \cite{De-Co:98} on the
asymptotics of orthogonal polynomials on the whole line with the weight $%
e^{-nV(\lambda )}$. In this paper we give another proof of the edge
universality of the eigenvalue statistics for the same class of potentials,
assuming additionally that these potentials lead to the DOS, whose support
is either an interval $[a,b]$ or, in the case of even potentials, that the
DOS\ support is $[-b,a]\cap \lbrack a,b]$, where $0<a<b<\infty $. The proof
is based on our recent results on the $1/n$-expansions for the matrix models
\cite{APS2}, establishing, in particular, the \textquotedblright slow
varying\textquotedblright\ character of the coefficients of the three-term
recurrent relation (the finite-difference equation) for respective
orthogonal polynomials. As a result, this relation becomes the Airy
differential equation, leading to the kernel (\ref{Aik}) in the interval (%
\ref{DeA}). We believe that our proof makes explicit an important
mathematical mechanism of the edge universality and is related to simplest
cases of the double scaling limit in the matrix models of the Quantum Field
Theory (see e.g. \cite{Bo-Br:91}, for the random matrix content of these
results).


\section{Main Result}

We will assume that the potential $V$, determining the probability law (\ref%
{p(M)}), satisfies the following conditions, in addition to conditions (\ref%
{log}) and (\ref{Lip}) above.

\medskip \noindent \textbf{Condition C1.} \textit{\ The support }$\sigma $
\textit{of the IDS of the ensemble consists of either }

\begin{enumerate}
\item[(i)] \textit{\ a single interval:
\begin{equation*}
\sigma =[a,b],\quad -\infty <a<b<\infty ,
\end{equation*}
or }

\item[(ii)] \textit{\ two symmetric intervals:
\begin{equation}
\sigma =[-b,-a]\cup \lbrack a,b],\quad 0<a<b<\infty ,  \label{sab}
\end{equation}
} \textit{and $V$ is even: $V(\lambda )=V(-\lambda )$, $\lambda \in \mathbf{R%
}$}.
\end{enumerate}

\noindent \textit{Remark}. It is easy to see that changing the variables
according to $M^{\prime }=M-\displaystyle\frac{a+b}{2}I$ in case (i) we can
always take the support $\sigma $ to be symmetric with respect to the
origin. Therefore without loss of generality we can assume that in this case
\begin{equation}
\sigma =[-a,a].  \label{saa}
\end{equation}

\medskip \noindent \textbf{Condition C2. }\textit{The DOS $\rho (\lambda )$
is strictly positive in the interior of the support $\sigma $ and behaves
asymptotically as $\,\mathrm{const}\cdot |\lambda -a_{\ast }|^{1/2}$, $%
\lambda \rightarrow a_{\ast }$, in a neighborhood of each edge }$a_{\ast }$%
\textit{\ of the support. Besides, the function
\begin{equation}
u(\lambda )=2\int \log |\mu -\lambda |\rho (\mu )d\mu -V(\lambda )  \label{u}
\end{equation}%
attains its maximum if and only if $\lambda $ belongs to the interior of the
closed set $\sigma $. We will call this behavior generic (see e.g. \cite%
{Kr-Mc:00} for results, justifying the term)}

\medskip \noindent \textbf{Condition C3. }\textit{$V(\lambda )$ is real
analytic on $\sigma $, i.e., there exists an open domain $\mathbf{D}\subset
\mathbf{C}$ and an analytic in $\mathbf{D}$ function $\mathcal{V}(z),z\in
\mathbf{D}$ such that}
\begin{equation*}
\sigma \subset \mathbf{D},\mathbf{\;}\mathcal{V}(\lambda +i0)=V(\lambda
),\quad \lambda \in \sigma .
\end{equation*}%
We mention that we always have the single interval case if $V$ is convex
\cite{BPS,Jo:98} or if it has a unique absolute minimum and sufficiently
large amplitude \cite{Kr-Mc:00}, and we always have the two interval case if
$V$ has two equal absolute minima and sufficiently large amplitude.
Conditions C2 and C3 were used in paper \cite{De-Co:99} to obtain asymptotic
formulas for orthogonal polynomials with the weight $e^{-nV}$. The condition
C3 is the case in many applications of the random matrix theory to the
Quantum Field Theory \cite{DiF:95} and to the condensed matter theory \cite%
{We-Co:98,Be}, where $V$ is often a polynomial.

The following statement known, in fact, in several contexts, provides a
sufficiently explicit form of the DOS in our case (see e.g \cite{APS2} for a
proof). 
\noindent

\begin{proposition}
\label{pro:0} Consider an ensemble of form (\ref{p(M)})--(\ref{dM}),
satisfying conditions (\ref{log}), and C1--C3 above. Then its density of
states (DOS) $\rho $ has the form
\begin{equation}
\rho (\lambda )=\frac{1}{2\pi }\chi _{\sigma }(\lambda )P(\lambda
)X_{+}(\lambda ),  \label{rho}
\end{equation}%
where $\chi _{\sigma }(\lambda )$ is the indicator of the support $\sigma $
of the DOS,
\begin{equation}
X_{+}(\lambda )=\left\{
\begin{array}{ll}
\sqrt{a^{2}-\lambda ^{2}}, & |\lambda |\leq a,\;\mathrm{in\;case}\;(\ref{saa}%
), \\
\mathrm{sign}\;\lambda \cdot \sqrt{(\lambda ^{2}-a^{2})(b^{2}-\lambda ^{2})},
& a\leq |\lambda |\leq b,\;\mathrm{in\;case}\;(\ref{sab}),%
\end{array}%
\right.  \label{X_+}
\end{equation}%
and
\begin{equation}
P(z)=\frac{1}{\pi }\int_{\sigma }{\frac{\mathcal{V}^{\prime }(z)-V^{\prime
}(\lambda )}{z-\lambda }}\frac{d\lambda }{X_{+}(\lambda )}.  \label{P}
\end{equation}%
Besides, the Stieltjes transform
\begin{equation}
g(z)=\int {\frac{\rho (\mu )d\mu }{z-\mu }},\quad \Im z\not=0,  \label{g(z)}
\end{equation}%
of the DOS for $z\in \mathbf{D}$ satisfies the quadratic equation
\begin{equation}
g^{2}(z)-\mathcal{V}^{\prime }(z)g(z)+\mathcal{Q}(z)=0,\quad z\in \mathbf{D},
\label{g,P}
\end{equation}%
where
\begin{equation}
\mathcal{Q}(z)=\int {\frac{\mathcal{V}^{\prime }(z)-V^{\prime }(\lambda )}{%
z-\lambda }\rho (\lambda )d\lambda }.  \label{Q}
\end{equation}
\end{proposition}

Denote by $p_{n}(\lambda _{1},...,\lambda _{n})$ the joint eigenvalue
probability density which we assume to be symmetric without loss of
generality. It is known that \cite{Me:91}
\begin{equation}
p_{n}(\lambda _{1},...\lambda _{n})=Q_{n}^{-1}\prod_{1\leq j<k\leq
n}(\lambda _{j}-\lambda _{k})^{2}\exp \left\{ -n\sum_{l=1}^{n}V(\lambda
_{l})\right\} ,  \label{p(la)}
\end{equation}%
where $Q_{n}$ is the respective normalization factor. Let
\begin{equation}
p_{l,n}(\lambda _{1},...,\lambda _{l})=\int p_{n}(\lambda _{1},...,\lambda
_{l},\lambda _{l+1},...\lambda _{n})d\lambda _{l+1}...d\lambda _{n}
\label{dp_k}
\end{equation}%
be the $l$th marginal distribution density of (\ref{p(la)}). Define the
correlation functions as
\begin{equation}
\mathcal{R}_{l,n}(\lambda _{1},...,\lambda _{l})=\frac{n!}{(n-l)!}%
p_{l}^{(n)}(\lambda _{1},...,\lambda _{l}).  \label{cf}
\end{equation}%
The link with orthogonal polynomials is provided by the formulas \cite%
{Me:91,BIZ}
\begin{equation}
\mathcal{R}_{l,n}(\lambda _{1},...,\lambda _{l})=\det \{K_{n}(\lambda
_{j},\lambda _{k})\}_{j,k=1}^{l},  \label{p_k=}
\end{equation}%
\begin{equation}
E_{n}(\Delta _{n})=\sum_{l=0}^{n-1}\frac{(-1)^{l}}{l!}\int_{\Delta _{n}^{l}}%
\hbox{det}\{K_{n}(\lambda _{j},\lambda _{k})\}_{j,k=1}^{l}d\lambda _{1}\dots
d\lambda _{l}.  \label{det}
\end{equation}%
Here
\begin{equation}
K_{n}(\lambda ,\mu )=\sum_{l=0}^{n-1}\psi _{l}^{(n)}(\lambda )\psi
_{l}^{(n)}(\mu )  \label{k_n}
\end{equation}%
is known as the reproducing kernel of the orthonormalized system,
\begin{equation}
\psi _{l}^{(n)}(\lambda )=\exp \{-nV(\lambda )/2\}p_{l}^{(n)}(\lambda
),\;\,l=0,...,  \label{psi}
\end{equation}%
in which $p_{l}^{(n)},\;l=0,...$ are orthogonal polynomials on $\mathbf{R}$
associated with the weight
\begin{equation}
w_{n}(\lambda )=e^{-nV(\lambda )},  \label{w}
\end{equation}%
i.e.,
\begin{equation}
\int p_{l}^{(n)}(\lambda )p_{m}^{(n)}(\lambda )w_{n}(\lambda )d\lambda
=\delta _{l,m}.  \label{ortP}
\end{equation}%
The polynomial $p_{l}^{(n)}$ has the degree $l$ and a positive coefficient $%
\gamma _{l}^{(n)}$ in front of $\lambda ^{l}$. The orthonormalized functions
$\{\psi _{l}^{(n)}\}_{l\geq 0}$ of (\ref{psi}) verify the recurrent
relations
\begin{equation}
J_{l}^{(n)}\psi _{l+1}^{(n)}(\lambda )+q_{l}^{(n)}\psi _{l}^{(n)}(\lambda
)+J_{l-1}^{(n)}\psi _{l-1}^{(n)}(\lambda )=\lambda \psi _{l}^{(n)}(\lambda
),\quad J_{-1}(n)=0,\quad l=0,....  \label{rec}
\end{equation}%
According to condition (\ref{log}) the polynomials $p_{l}^{(n)}$ and the
coefficients $J_{l}^{(n)}$ are defined for all $l$ such that
\begin{equation}
l\leq n_{1},\quad n_{1}=n(1+\epsilon /4).  \label{n_1}
\end{equation}%
In other words, we have here the $n_{1}\times n_{1}$ real symmetric Jacobi
matrix
\begin{equation}
\begin{array}{l}
J^{(n)}=\{J_{l,m}^{(n)}\}_{l,m=0}^{n_{1}}, \\
J_{l,m}^{(n)}=q_{l}^{(n)}\delta _{l,m}+J_{l}^{(n)}\delta
_{l+1,m}+J_{l-1}^{(n)}\delta _{l-1,m}.%
\end{array}
\label{J}
\end{equation}%
Note that if $V$ is even, then $q_{l}^{(n)}=0$, $l=0,\dots $.

We will need an important particular case of the above formulas \cite{Me:91}%
, corresponding to $l=1$ in (\ref{p_k=}):

\begin{equation}
\mathbf{E}_{n}\{{N_{n}(\Delta )\}}=\int_{\Delta }\rho _{n}(\lambda )d\lambda
,\;\;\rho _{n}(\lambda )=n^{-1}K_{n}(\lambda ,\lambda ),  \label{Nbar}
\end{equation}
where the symbol $\mathbf{E}_{n}\{{...\}}$ denotes the expectation with
respect to the measure, defined by (\ref{p(la)}), i.e., by (\ref{p(M)}) - (%
\ref{dM}).

\noindent

\begin{theorem}
\label{thm:1} Consider an ensemble of the form (\ref{p(M)}) - (\ref{dM}),
satisfying conditions (\ref{Lip}), (\ref{log}), and C1 - C3 above. Then for
any  endpoint $a_{\ast }$ of the support $\sigma $ and for any
positive integer $l$ the rescaled correlation function
\begin{equation}
(\gamma n^{2/3})^{-l}\mathcal{R}_{l,n}(a_{\ast }\pm t_{1}/\gamma
n^{2/3},...,a_{\ast }\pm t_{l}/\gamma n^{2/3})  \label{cfresc}
\end{equation}%
converges weakly as $n\rightarrow \infty $ to
\begin{equation}
\det \{\mathcal{K}(t_{j},t_{k})\}_{j,k=1}^{l},  \label{det1}
\end{equation}%
where the sign $\pm$ (\ref{cfresc}) corresponds to a right hand and left hand
endpoint, $\mathcal{K}(t_{j},t_{k})$ is the Airy kernel (\ref{Aik}),
\begin{equation}
\gamma =(2c^2\alpha)^{-1/3}  \label{gamma}
\end{equation}%
and
\begin{equation} \alpha=a,\quad
c={{\frac{1}{2a}}\bigg({\frac{1}{P(a)}}+{\frac{1}{P(-a)}}\bigg)}  \label{ci}
\end{equation}%
in the case (\ref{saa}) and
\begin{equation}
\alpha=(b^2-a^2)\left\{
\begin{array}{ll} a^{-1}, &  \\
b^{-1}, &
\end{array}\right.
c=\left\{
\begin{array}{ll}
\displaystyle\frac{2}{(b^{2}-a^{2})P(a)}{,} &  \\
\displaystyle\frac{2}{(b^{2}-a^{2})P(b)}, &
\end{array}%
\right.  \label{cii}
\end{equation}%
in the case (\ref{sab}) for the endpoint $a$ and $b$ respectively. The
function $P(\lambda )$, entering (\ref{ci}) and (\ref{cii}) is defined in (%
\ref{P}).

Besides, if $\Delta \subset \mathbf{R}$ is a finite union of disjoint
intervals bounded from the left in the case of the right hand endpoint
 and from the right in the case of the left hand endpoint, and $E_{n}(\Delta )$ is the hole probability
(\ref{hon}) for $\Delta _{n}=a_*\pm \Delta /\gamma n^{2/3}$, then
\begin{equation}
\lim_{n\rightarrow \infty }E_{n}(\Delta _{n})=1+\sum_{l=1}^{\infty }\frac{%
(-1)^{l}}{l!}\int_{\Delta }dt_{1}\dots dt_{l}\det \{\mathcal{K}%
(t_{j},t_{k})\}_{j,k=1}^{l},  \label{t1.2}
\end{equation}%
i.e., the limit is the Fredholm determinant of the integral operator $%
\mathcal{K}_{\Delta }$, defined by the kernel (\ref{Aik}) on the set $\Delta
$ and the sign $\pm $ in $\Delta _{n}$ corresponds to the right and left
hand endpoints of the support.
\end{theorem}

We mention now two particular cases of the theorem. The first corresponds to
the case $l=1$. 
\noindent

\begin{corollary}
\label{cor.1} Denote
\begin{equation}
\nu _{n}(s)=\rho _{n}(a_{\ast }\pm s/\gamma n^{2/3})n^{1/3}/\gamma .
\label{nuns}
\end{equation}%
Then we have weakly in $\mathbf{R:}$%
\begin{equation*}
\lim_{n\rightarrow \infty }\nu _{n}=\nu ,
\end{equation*}%
where
\begin{equation}
\nu (s)=\int_{s}^{\infty }Ai^{2}(x)dx,  \label{nuAi}
\end{equation}%
and $\pm $ in (\ref{nuns}) corresponds to the right and left hand endpoints
respectively.
\end{corollary}

\noindent \textit{Remarks}. 1). Denote $\mathcal{N}_{n}(\Delta )$ the number
of eigenvalues in the set $\Delta _{n}=a_{\ast }\pm \Delta /\gamma
n^{2/3},\;\Delta \subset \mathbf{R}$. According to (\ref{Nbar}) and (\ref%
{nuns})
\begin{equation*}
\mathbf{E}_{n}\{\mathcal{N}_{n}(\Delta )\}=n\int_{a_{\ast }+2cn^{-2/3}\Delta
}\rho _{n}(\lambda )d\lambda =\int_{\Delta }\nu _{n}(s)ds.
\end{equation*}%
Hence, we can interpret the corollary as a statement, according to which the
expectation of the rescaled counting measure $\mathcal{N}_{n}$ converges
weakly to the absolute continuous measure $\mathcal{N}$ whose density is (%
\ref{nuAi}). The density can be viewed as an analogue of the density of
states for $n^{-2/3}$- neighborhoods of the edge $a_{\ast }$ of the support
of the Density of States, given by Proposition \ref{pro:0}.

2). By using the equation%
\begin{equation}
Ai^{^{\prime \prime }}(x)=xAi(x),  \label{Aie}
\end{equation}%
and the following from the equation identity \cite{AS}
\begin{equation}
\frac{Ai(x)Ai^{\prime }(y)-Ai^{\prime }(x)Ai(y)}{x-y}=\int_{0}^{\infty
}Ai(x+u)Ai(y+u)du  \label{AiCD}
\end{equation}%
for $x=y$, we can rewrite the r.h.s. of (\ref{nuAi}) in the form
\begin{equation}
\nu (s)=Ai^{\prime 2}(s)-sAi^{2}(s).  \label{nuFo}
\end{equation}%
The formula was obtained in \cite{Fo:93} in the case (\ref{Ga}) of the GUE,
by using (\ref{Nbar}) and the Plancherel-Rotah asymptotic formula for the
Hermite polynomials, that play the role of polynomials $p_{l}^{(n)}$ for the
GUE \cite{Me:91}.

The next corollary deals with the case of $\Delta =(s,\infty )$ of formula (%
\ref{t1.2}).

\noindent

\begin{corollary}
Under condition of Theorem \ref{thm:1}\ the $n=\infty $ limit of the
probability distribution of the maximum eigenvalue of the random matrix (\ref%
{p(M)}) - (\ref{dM}) is
\begin{equation*}
\lim_{n\rightarrow \infty }\mathbf{P}_{n}\{\lambda _{\max }^{(n)}\leq
a_{\ast }+s/\gamma n^{2/3}\}=\det (1-\mathcal{K}(s)),
\end{equation*}%
where $a_{\ast }$ is  the extreme  right-hand endpoint of the support and $\mathcal{K}%
(s) $ is the integral operator, defined by the kernel (\ref{Aik}) on the
interval $(s,\infty )$.
\end{corollary}

The corollary asserts that the $n=\infty $ limit of the probability
distribution of the maximum eigenvalue of the random matrix is independent
of the ensemble (of function $V$ in \ref{p(M)}), i.e., the universality of
this distribution for the class of ensembles, treated in the paper.
Analogous statement is valid for the minimum eigenvalue.

\section{Proof of the Main Result}

We will prove Theorem \ref{thm:1} in details for the case (i), where the
support of the Density of States is an interval of the spectral axis. At the
end of the proof we shall explain the difference between this case and the
two-interval case (ii). Besides, we can restrict ourselves to the right hand
endpoint $a$ of the interval $[-a,a]$ without loss of generality.

The proof is based on the following asymptotic formula for the coefficient $%
J_{l}^{(n)}$ of the Jacobi matrix (\ref{J}) \cite{APS2}, Theorem 1:
\begin{equation}
J_{n+k}^{(n)}={\frac{a}{2}}+{\frac{k}{n}}c+r_{k}^{(n)},\quad q_{n+k}=%
\overline{r}_{k}^{(n)},  \label{Ji}
\end{equation}%
where  $c$ is given by (\ref{ci}), and the remainders $r_{k}^{(n)}$,
$\overline{r}%
_{k}^{(n)}$ admit the estimate
\begin{equation}
|r_{k}^{(n)}|,|\overline{r}_{k}^{(n)}|\leq \,C\cdot {\frac{k^{2}+1}{n^{2}}}%
,\,\ |k|\leq C\cdot n^{2/3}.  \label{r_k}
\end{equation}%
Here and below the symbol $C$ denotes positive quantities that do not depend
on $n$ and $k$ but can be different in different formulas.

\medskip

It follows from (\ref{rec}) and the orthonormality of $\{p_{l}^{(n)}\}_{l%
\geq 0}$ that
\begin{equation}
J_{l}^{(n)}=\int \lambda p_{l+1}^{(n)}(\lambda )p_{l}^{(n)}(\lambda
)d\lambda .  \label{Jpp}
\end{equation}%
This and (\ref{Ji}) imply that the order of the orthogonal polynomials $%
p_{l}^{(n)}$, entering formulas (\ref{p_k=}) - (\ref{k_n}), and (\ref{Ji})
does not exceed $n+Cn^{2/3}$, and makes possible to replace $p_{l}^{(n)}$ of
(\ref{ortP}), orthogonal on the whole axis, by the polynomials $%
p_{l}^{(L,n)} $, orthogonal on a sufficiently big but finite interval $%
[-L,L] $ with respect to the same weight (\ref{w}). This will simplify our
analysis and is justified by the following

\begin{lemma}
\label{lem:*} Let $\{p_{l}^{(L,n)}\}_{l=0}^{\infty }$ is the system of
polynomials orthogonal on the interval $[-L,L]$ with respect to the weight (%
\ref{w}):
\begin{equation*}
\int_{L}^{L}p_{l}^{(L,n)}(\lambda )p_{m}^{(L,n)}(\lambda )e^{-nV(\lambda
)}=\delta _{l,m}.
\end{equation*}%
Denote by $\psi _{l}^{(L,n)}$, $K_{n}^{(L)}$, and $J_{l}^{(L,n)}(n)$ the
quantities defined in (\ref{psi}), (\ref{k_n}), and (\ref{rec}) for the
system $\{p_{l}^{(L,n)}(\lambda )\}_{l=0}^{\infty }$. Assume, that $%
V(\lambda )$ satisfies conditions (\ref{log}) and (\ref{Lip}). Then there
exist some absolute constants $L$ and $L_{1}$ such that for any $0\leq l\leq
(1+\epsilon /4)n$
\begin{equation}
\max_{|\lambda |\leq L}\bigg|\psi _{l}(\lambda )-\psi _{l}^{(L)}(\lambda )%
\bigg|\leq C\,e^{-nL_{1}},\quad (\lambda \in \lbrack -L,L]).  \label{l*.1}
\end{equation}%
\begin{equation}
|J_{l}^{(L,n)}-J_{l}^{(n)}|\leq C\,e^{-nL_{1}}  \label{JL}
\end{equation}%
\begin{equation}
\max_{|\lambda |,|\mu |\leq L|}|K_{n}(\lambda ,\mu )-K_{n}^{(L)}(\lambda
,\mu )|\leq C\,e^{-nL_{1}}  \label{KL}
\end{equation}
\end{lemma}

The lemma allows us to substitute $R_{l}^{(n)}$, $E_{n}(\Delta _{n})$ and $%
J_{l}^{(n)}$ in (\ref{cfresc}), (\ref{t1.2}) and (\ref{Ji}) by the
respective quantities, constructed from the polynomials $\{p_{l}^{(L,n)}%
\}_{l=0}^{\infty }$. We will assume from now on that this replacement is
made and we will omit the super index $(L)$ to simplify notations.

\medskip

Now we will prove the first assertion of the theorem, relations (\ref{cfresc}%
) - (\ref{cii}). Since any permutation of $l$ objects can be represented as
the product of the cyclic permutations, each term of the determinant in (\ref%
{p_k=}) is the product of the expressions ${K_{n}}\left( \lambda _{1}{%
,\lambda _{2}}\right) {\dots {K_{n}}\left( \lambda _{m-2}{,\lambda _{m-1}}%
\right) {K_{n}(}\lambda _{m-1}{,\lambda _{1})}}$, in which the arguments are
in the cyclic order and do not appear in any other cyclic expression of the
product. Hence, it suffices to prove the weak limit%
\begin{equation}
\begin{array}{r}
\lim_{n\rightarrow \infty }(\gamma {n^{2l/3})}^{-l}{K_{n}}\left( {a+{t}%
_{1}/\gamma n}^{2/3}{,a+{t}_{2}/\gamma n}^{2/3}\right) {\dots K_{n}}\left( {%
a+t}_{m}{/\gamma n}^{2/3}{,a+{t}_{1}/\gamma n}^{2/3}\right) \\
={\mathcal{K}(t_{1},t_{2})\dots \mathcal{K}(t_{m},t_{1})}%
\end{array}
\label{final}
\end{equation}%
for any $m\geq 1$. We will confine ourselves to the case $m=2$, containing
all important ingredients of the general case.

We set $z_{1}=a+n^{-2/3}\zeta _{1}$, $z_{2}=a+n^{-2/3}\zeta _{2}$ and
introduce the function
\begin{equation}
F_{n}(\zeta _{1},\zeta _{2})=n^{-4/3}\int \int \Im \frac{1}{\lambda
_{1}-z_{1}}{\Im \frac{1}{(\lambda _{2}-z_{2})}}K_{n}^{2}(\lambda
_{1},\lambda _{2})d\lambda _{1}d\lambda _{2}  \label{K^2(x,y)}
\end{equation}%
for $|\Im \zeta _{1,2}|\geq \varepsilon _{0}>0$, i.e., the two-dimensional
Poisson integral of the function $K_{n}^{2}(\lambda _{1},\lambda _{2})$.
According to (\ref{psi}) - (\ref{rec}) the functions $\{\psi
_{l}^{(n)}\}_{l=0}^{\infty }$ are the generalized eigenfunctions of the
selfadjoint operator, defined in the space $l^{2}(\mathbf{Z}_{+})$ by the
matrix $J^{(n)}$ of (\ref{J}). Denoting the operator again $J^{(n)}$, we
introduce its resolvent
\begin{equation}
R^{(n)}(z)=(J^{(n)}-zI)^{-1},  \label{resJ}
\end{equation}%
and the matrix $\{R_{j,k}^{(n)}(z)\}_{j,k=0}^{\infty }$ of the resolvent in
the canonical basis of $l^{2}(\mathbf{Z}_{+})$. Then the spectral theorem
yields the representation
\begin{equation}
R_{j,k}^{(n)}(z)=\int \frac{\psi _{j}^{(n)}(\lambda )\psi _{k}^{(n)}(\lambda
)}{\lambda -z}d\lambda ,\quad \Im z\neq 0.  \label{Rjk}
\end{equation}%
By using (\ref{k_n}) and this representation we obtain that the function $%
F_{n}$ of (\ref{K^2(x,y)}) can be written as follows:
\begin{equation}
F_{n}(\zeta _{1},\zeta _{2})=n^{-4/3}\sum_{j,k=1}^{n}\Im
R_{n-j,n-k}^{(n)}(a+n^{-2/3}\zeta _{1})\Im R_{n-j,n-k}^{(n)}(a+n^{-2/3}\zeta
_{2}).  \label{FImR}
\end{equation}%
Set
\begin{equation}
M=[n^{3/5}],  \label{M}
\end{equation}%
\noindent

\begin{lemma}
\label{lem:1} Let $R^{(n)}(z)$ be the resolvent (\ref{resJ}) of $J^{(n)}$.
Then for any $z=a+n^{-2/3}\zeta$,  $|\Im \zeta|>e^{-{C\,\sqrt n}}$, $\Re
\zeta\ge -C$ we have
\begin{equation}
n^{-4/3}\sum_{j=M+1}^{n}\sum_{k=0}^{\infty }|R^{(n)}_{n-j,k}(z)|^{2}\leq C\,
\frac{n^{5/3}}{M^{3}}=C\, n^{-2/15}.  \label{est2}
\end{equation}
\end{lemma}

The proof of the lemma will be given in the next section.

Consider the operator
\begin{equation}
{A}=\displaystyle{{\frac{a}{2}}{\frac{d^{2}}{dx^{2}}}-2cx}  \label{Aop}
\end{equation}%
defined on the whole real line. Denote by ${R}(\zeta )$ the resolvent $({A}%
-\zeta I)^{-1}$ of ${A}$ for $\Im \zeta \neq 0$, and by $R_{\zeta }(x,y)$
the kernel of ${R}(\zeta )$. We will need the following

\noindent

\begin{proposition}
\label{pro:d1} The kernel $R_{\zeta }(x,y)$ possesses the properties:

\begin{enumerate}
\item[(i)]
\begin{equation}
\frac{a}{2}\frac{\partial ^{2}}{\partial x^{2}}R_{\zeta }(x,y)-2cxR_{\zeta
}(x,y)=\zeta R_{\zeta }(x,y)+\delta (x-y),  \label{eqRz}
\end{equation}

\item[(ii)]
\begin{equation}
R_{\zeta }(x,y)=\frac{2\pi }{\kappa a}\left\{
\begin{array}{ll}
\psi _{-}(x,\zeta )\psi _{+}(y,\zeta ) & x\leq y, \\
\psi _{+}(x,\zeta )\psi _{-}(y,\zeta ) & x\geq y,%
\end{array}%
\right.  \label{R-psi}
\end{equation}%
where
\begin{equation}
\psi _{+}(x,\zeta )=\emph{\hbox{Ai}}(X),\quad \psi _{-}(x,\zeta )=\emph{%
\hbox{Ci}}(X),\quad  \label{psi_pm}
\end{equation}%
\begin{equation}
\emph{\hbox{Ci}}(X)=i\emph{\hbox{Ai}}(X)-\emph{\hbox{Bi}}(X)  \label{Ci}
\end{equation}%
$X=\kappa x+\gamma \zeta $, $\kappa =(4ca^{-1})^{1/3}$, and $\emph{\hbox{Ai}}%
(z)$ and $\emph{\hbox{Bi}}(z)$ are the standard Airy functions (see \cite{AS}%
).

\item[(iii)] The functions $\psi _{\pm }$ are entire in $x$ and $\zeta $ and
have the following asymptotic behavior in $x$ for $\Im \zeta =\varepsilon
>0$
\begin{equation}
|\psi _{+}(x,\zeta )|=\displaystyle\frac{1}{\pi ^{1/2}|X|^{1/4}}%
(1+O(|x|^{-3/2})\left\{
\begin{array}{ll}
\exp \{-\displaystyle\frac{2}{3}|\Re X|^{3/2}\}, & x\rightarrow +\infty \\
\exp \{\gamma \varepsilon |\Re X|^{1/2}\}, & x\rightarrow -\infty%
\end{array}%
\right.  \label{as_psi+}
\end{equation}%
\begin{equation}
|\psi _{-}(x,\zeta )|=\displaystyle\frac{1}{2\pi ^{1/2}|X|^{1/4}}%
(1+O(|x|^{-3/2})\left\{
\begin{array}{ll}
\exp \{\displaystyle\frac{2}{3}|\Re X|^{3/2}\}, & x\rightarrow +\infty \\
\exp \{-\gamma \varepsilon |\Re X|^{1/2}\}, & x\rightarrow -\infty%
\end{array}%
\right.  \label{as_psi-}
\end{equation}

\item[(iv)] if $I(x,y)=\Im R_{\zeta }(x,y)$, then
\begin{equation}
|I(x,y)|^{2}\leq I(x,x)I(y,y),\;\mathrm{and\;}\int_{-\infty
}^{0}I(x,x)dx<\infty .  \label{I}
\end{equation}
\end{enumerate}
\end{proposition}

The proposition will be proved in the next section.

Introduce the double infinite matrix
\begin{equation}
R_{l_{1},l_{2}}^{\ast }(z)=n^{1/3}R_{\zeta }\left( {\frac{n-l_{1}}{n^{1/3}}},%
{\frac{n-l_{2}}{n^{1/3}}}\right) ,\quad z=a+n^{-2/3}\zeta ,  \label{R}
\end{equation}%
and the semi infinite matrix
\begin{equation}
D=\{d_{l_{1},l_{2}}\}_{l_{1},l_{2}=0}^{\infty },\;D=(J^{(n)}-zI)R^{\ast
}(z)-I.  \label{D}
\end{equation}%
Then we have
\begin{equation}
R^{(n)}(z)=R^{\ast }(z)-R^{(n)}D(z).  \label{RRD}
\end{equation}%
We introduce also the $(4M+1)\times (4M+1)$ matrix $D^{(M)}$, assuming that $%
n$ is big enough and setting
\begin{equation}\label{ti-D}
D^{(M)}=\{D_{l_{1},l_{2}}^{(M)}\},\;D_{n-j,n-k}^{(M)}=\left\{
\begin{array}{cc}
d_{n-j,n-k}, & |j|,|k|\leq 2M, \\
0, & \mathrm{otherwise.}%
\end{array}%
\right.
\end{equation}%
We will use the following lemmas, that will also be proved in the next
section.

\noindent

\begin{lemma}
\label{lem:2} Under conditions of Theorem \ref{thm:1} for any $%
z=a+n^{-2/3}\zeta $ with $|\Re \zeta |\leq C$ and $|\Im \zeta |\geq
\varepsilon =O(n^{-\alpha _{1}})$, $0\leq \alpha _{1}\leq 1/11,\;\alpha
_{2}\geq 0$ we have%
\begin{equation}
n^{-4/3}\sum_{j=1}^{M}\sum_{|k|\leq M}|R_{n-j,n-k}^{(n)}(z)|^{2}\leq
2n^{-4/3}\sum_{j=1}^{M}\sum_{|k|\leq 2M}|R_{n-j,n-k}^{\ast
}(z)|^{2}+CM/n^{5/2}+O(n^{-1}),  \label{est3}
\end{equation}%
\begin{equation}
n^{-4/3}\sum_{j=1}^{M}\sum_{|k|\leq M}|(R^{(n)}D)_{n-j,n-k}(z)|^{2}\leq
C\varepsilon ^{-1}||D^{(M)}||^{2}.  \label{est3a}
\end{equation}
\end{lemma}

\begin{lemma}
\label{lem:D} If $|\Im \zeta |\geq \varepsilon >0$, then the norm of the matrix $%
D^{(M)}$ admits the bound%
\begin{equation}
||D^{(M)}||\leq C\left( (M/n)^{1/2}+\varepsilon ^{-1}M^{2}/n^{4/3}\right) .
\label{||D||}
\end{equation}
\end{lemma}

\bigskip

On the basis of (\ref{est2}) and (\ref{RRD}),  we get for (\ref{FImR})
\begin{equation}
\begin{array}{c}
\displaystyle{F(\zeta _{1},\zeta _{2})=n^{-4/3}\sum_{j,k=1}^{M}\Im
R_{n-j,n-k}^{(n)}(z_{1})\Im R_{n-j,n-k}^{(n)}(z_{2})+O(n^{-2/15})} \\
\displaystyle{=n^{-4/3}\sum_{j,k=1}^{M}\Im R_{n-j,n-k}^{\ast }(z_{1})\Im
R_{n-j,n-k}^{\ast }(z_{2})+\delta _{n}(\zeta _{1},\zeta _{2})+O(n^{-2/15})},%
\end{array}
\label{main}
\end{equation}%
where
\begin{equation*}
\begin{array}{c}
|\delta _{n}(\zeta _{1},\zeta _{2})|=\bigg|n^{-4/3}\displaystyle%
\sum_{j,k=1}^{M}\bigg[\Im (R^{(n)}D)_{n-j,n-k}(z_{1})\Im R_{n-j,n-k}^{\ast
}(z_{2})+ \\
\Im (R^{(n)}D)_{n-j,n-k}(z_{2})\Im R_{n-j,n-k}^{\ast }(z_{1})+\Im
(R^{(n)}D)_{n-j,n-k}(z_{1})\Im (R^{(n)}D)_{n-j,n-k}(z_{2})\bigg| \\
\displaystyle{\leq \bigg[n^{-4/3}%
\sum_{j,k=1}^{M}|(R^{(n)}D)_{n-j,n-k}(z_{1})|^{2}\bigg]^{1/2}\bigg[%
n^{-4/3}\sum_{j,k=1}^{M}|R_{n-j,n-k}^{\ast }(z_{2})|^{2}\bigg]^{1/2}} \\
\displaystyle{+\bigg[n^{-4/3}%
\sum_{j,k=1}^{M}|(R^{(n)}D)_{n-j,n-k}(z_{2})|^{2}\bigg]^{1/2}\bigg[%
n^{-4/3}\sum_{j,k=1}^{M}|R_{n-j,n-k}^{\ast }(z_{1})|^{2}\bigg]^{1/2}} \\
\displaystyle{+\bigg[n^{-4/3}%
\sum_{j,k=1}^{M}|(R^{(n)}D)_{n-j,n-k}(z_{1})|^{2}\bigg]^{1/2}\bigg[%
n^{-4/3}\sum_{j,k=1}^{M}|(R^{(n)}D)_{n-j,n-k}(z_{2})|^{2}\bigg]^{1/2}}%
\end{array}%
\end{equation*}%
Since we have to prove the convergence of $F_{n}$ for $\zeta
_{1,2}$ such that $|\zeta _{1,2}|\leq C$ and $|\Im \zeta
_{1,2}|\geq \varepsilon _{0}>0$,
we can take \ an $n$-independent $\varepsilon =\varepsilon _{0}$ in Lemmas %
\ref{lem:2}, and \ref{lem:D}. Then, by using the Euler-MacLaurin
summation formula \cite{AS}, it can be shown that the sum in the l.h.s. of (\ref{est3})
is $O(1)$ as $n\rightarrow \infty $. This leads to the bound
\begin{equation*}
|\delta _{n}(\zeta _{1},\zeta _{2})|\leq C\,n^{-2/15}.
\end{equation*}%
The bound, Proposition \ref{pro:d1}, (\ref{main}), and (\ref{R}) imply that
for any $\zeta _{1,2}$ such that $|\zeta _{1,2}|\leq C<\infty $ and $|\Im
\zeta _{1,2}|\geq \varepsilon _{0}>0$, we have
\begin{equation}
\lim_{n\rightarrow \infty }{F}_{n}(\zeta _{1},\zeta _{2})=\int_{0}^{\infty
}\int_{0}^{\infty }\Im R_{\zeta _{1}}(x_{1},x_{2})\Im R_{\zeta
_{2}}(x_{1},x_{2})dx_{1}dx_{2}.  \label{lFn}
\end{equation}%
According to (\ref{R-psi}) the function $(2/\kappa a)^{1/2}Ai(\kappa
x+\gamma \xi )$ is the generalized eigenfunction of the self-adjoint
operator (\ref{Aop}), corresponding to the generalized eigenvalue $\xi \in
\mathbf{R}$. This fact and the spectral theorem for the operator yield the
integral representation
\begin{equation}
\Im R_{\zeta }(x,y)=\frac{2}{\kappa a}\int Ai(\kappa x+\gamma \xi )Ai(\kappa
y+\gamma \xi )\Im \frac{1}{\xi -\zeta }d\xi .  \label{spAi}
\end{equation}%
The formula and Proposition \ref{pro:d1} allow us to rewrite the r.h.s. \ of
(\ref{lFn}) as
\begin{eqnarray*}
&&\gamma ^{2}\int_{0}^{\infty }\int_{0}^{\infty }\Im \frac{1}{\xi _{1}-\zeta
_{1}}\Im \frac{1}{\xi _{2}-\zeta _{2}}d\xi _{1}d\xi _{2} \\
&&\times \int_{0}^{\infty }\int_{0}^{\infty }dx_{1}dx_{2}\hbox{Ai}%
(x_{1}+\gamma {\xi }_{1})\hbox{Ai}(x_{1}+\gamma \xi _{2})\hbox{Ai}({x}%
_{2}+\gamma \xi _{1})\hbox{Ai}({x}_{2}+\gamma \xi _{2}).
\end{eqnarray*}%
To finish the proof we need the following lemma, proved in the next section.

\begin{lemma}
\label{pro:3} Under conditions of Theorem \ref{thm:1} for any fixed $%
L_{0}>-\infty $
\begin{equation}
n\int_{\mathbf{R}\setminus \sigma (n^{-2/3}L_{0})}\rho _{n}(\lambda
)d\lambda \leq C\,\int_{L_{0}}^{\infty }dt\int_{0}^{\infty }dx\emph{\hbox{Ai}%
}^{2}(\kappa x+\gamma t)+o(1),\quad n\rightarrow \infty ,  \label{teil_rho}
\end{equation}%
where $\sigma (\varepsilon )$ for $\varepsilon >0$ is the $\varepsilon $%
-neighborhood of $\sigma $ and for $\varepsilon <0\;\sigma (\varepsilon )$
denotes the part of $\sigma $, whose distance from the boundary of $\sigma $
exceeds $|\varepsilon |$.
\end{lemma}

Since the l.h.s. of (\ref{lFn}) is bounded from above for $|\Im \zeta|>0$, we conclude that the
sequence of measures in $\mathbf{R}^{2}$, defined by the densities $%
n^{-4/3}K_{n}^{2}(a+n^{-2/3}\xi _{1},a+n^{-2/3}\xi _{2})$ (cf (\ref{nuns})),
is weakly compact. Besides, Proposition \ref{pro:3} and the inequality
\begin{equation}
K_{n}^{2}(\lambda ,\mu )\leq K_{n}(\lambda ,\lambda )K_{n}(\mu ,\mu
)=n^{2}\rho _{n}(\lambda )\rho _{n}(\mu )  \label{Krr}
\end{equation}%
imply that the contributions of neighborhoods of $\pm \infty $ with respect
to the both variables in (\ref{K^2(x,y)}) is negligible uniformly in $n$.
Since the Poisson integral determines uniquely the  corresponding measure, we
deduce from the above and formulas (\ref{AiCD}), (\ref{FImR}), and (\ref{lFn}) \ the tight
convergence of $n^{-4/3}K_{n}^{2}(a+n^{-2/3}\xi _{1},a+n^{-2/3}\xi _{2})d\xi
_{1}d\xi _{2}$ to $\gamma ^{2}\mathcal{K}^{2}(\gamma\xi _{1},\gamma\xi _{2})d\xi
_{1}d\xi _{2}$. Changing variables $\xi _{1,2}$ to $t_{1,2}/\gamma $, we
obtain (\ref{cfresc}) - (\ref{det1}) for $l=2$.

This finishes the proof of the first assertion of the theorem, i.e. the
relations (\ref{cfresc})--(\ref{ci}), for the single interval support (\ref%
{saa}) of the Density of States of the ensemble. Let us prove now
the second assertion of the theorem, formula (\ref{t1.2}). We note
first that if $\Delta _{n}=a+\Delta /\gamma n^{2/3}$, then we have
according to (\ref{det})
\begin{equation}
\begin{array}{c}
E_{n}(\Delta _{n})=1+\displaystyle\sum_{l=1}^{n}\frac{(-1)^{l}}{l!}%
\displaystyle\int_{\Delta }dt_{1}\dots dt_{l}(\gamma n^{2/3})^{-l}\det %
\bigg\{K_{n}(a+{t}_{j}/\gamma n^{2/3},a+t_{k}/\gamma n^{2/3})\bigg\}%
_{j,k=1}^{l},%
\end{array}
\label{t1.3}
\end{equation}%
Recall now the Hadamard inequality, according to which we have for any $%
l\times l$ matrix $A=\{A_{jk}\}_{j,k=1}^{l}$:
\begin{equation*}
|\det A|\leq \prod_{j=1}^{l}\left( \sum_{k=1}^{n}|A_{jk}|^{2}\right) ^{1/2}.
\end{equation*}%
If the matrix is positive definite, the inequality can be modified as
follows
\begin{equation}
\det A\leq \prod_{j=1}^{l}A_{jj}.  \label{Hmod}
\end{equation}%
The last inequality and Proposition \ref{pro:3} allow us to make the limit $%
n\rightarrow \infty $ in the r.h.s. of (\ref{t1.2}), hence to prove (\ref%
{t1.2}) for any set $\Delta \subset \mathbf{R}$, finite or bounded from the
left.

This finishes the proof of Theorem \ref{thm:1} for the case (\ref{saa}) of a
single interval support of the Density of States of the ensemble and the right
hand endpoint $a$ of the support. The case of the left hand endpoint $-a$
can be treated analogously by setting $z=-a-n^{-2/3}\zeta$ and by using
$$
-n^{1/3}(-1)^{l_1+l_2}(A-\zeta)^{-1}
\bigg(\frac{n-l_1}{n^{1/3}},\frac{n-l_1}{n^{1/3}}\bigg)
$$
as the matrix $R^*$.

 To prove
the theorem for the case (\ref{sab}) of a two-interval support we note that
now  we
have the following asymptotic relation \cite{APS} (cf (\ref{Ji})):
\begin{equation}
J_{n+k}^{(n)}=\frac{1}{2}(b-(-1)^{k}a)+\frac{k}{n(b^{2}-a^{2})}\bigg(\frac{1%
}{P(b)}-\frac{(-1)^k}{P(a)}\bigg)+r_{k}^{(n)}  \label{Jii}
\end{equation}%
instead of (\ref{J}), where $r_{k}^{(n)}$ again satisfies (\ref{r_k}). As a
result, in the case the endpoint $b$ we should consider instead (\ref{Aop})
the operator
\begin{equation*}
{A}^{(b)}=\frac{b^{2}-a^{2}}{2b}{\frac{d^{2}}{dx^{2}}}-\frac{2}{%
(b^{2}-a^{2})P(b)}x
\end{equation*}%
with the resolvent ${R}^{(b)}(\zeta )=({A}^{(b)}-\zeta I)^{-1}$ whose kernel
is $R_{\zeta }^{(b)}(x,y)$. Then we consider the matrix $R^{(\ast ,b)}(\zeta
)$ of the form
\begin{equation}
R_{n-j,n-k}^{(\ast ,b)}(\zeta )=n^{1/3}R_{\zeta }^{(b)}\bigg(\frac{j}{n^{1/3}%
}+\frac{(-1)^{j}a}{2n^{1/3}b},\frac{k}{n^{1/3}}+\frac{(-1)^{k}a}{2n^{1/3}b}%
\bigg).  \label{Rii,b}
\end{equation}%
The respective operator for the endpoint $a$ is
\begin{equation*}
{A}^{(a)}=\frac{b^{2}-a^{2}}{2a}{\frac{d^{2}}{dx^{2}}}+\frac{2}{%
(b^{2}-a^{2})P(a)}x
\end{equation*}%
with the resolvent ${R}^{(a)}(\zeta )$ and
\begin{equation}
R_{n-j,n-k}^{(\ast ,a)}(\zeta )=n^{1/3}(-1)^{[\frac{k+1}{2}]}(-1)^{[\frac{j+1%
}{2}]}R_{\zeta }^{(a)}\bigg(\frac{j}{n^{1/3}}+\frac{(-1)^{j}b}{2n^{1/3}a},%
\frac{k}{n^{1/3}}+\frac{(-1)^{k}b}{2n^{1/3}a}\bigg)  \label{Rii,a}
\end{equation}%
where $R_{\zeta }^{(a)}(x,y)$ be the kernel of ${R}^{(a)}(\zeta )$.

Theorem \ref{thm:1} is proved.


\section{Auxiliary results}

\noindent \textit{Proof of Lemma \ref{lem:*}} We prove first (\ref{l*.1}).
For $\overline{\mu }=(\mu _{1},\dots ,\mu _{k})$ we set:
\begin{equation*}
\begin{array}{c}
\Pi _{k}(\overline{\mu })=\displaystyle\prod_{1\leq i<j\leq k}(\mu _{i}-\mu
_{j})^{2}\displaystyle\prod_{i=1}^{k}e^{-nV(\mu _{i})},\quad \Pi _{k}^{\ast
}(\lambda ,\overline{\mu })=e^{-nV(\lambda )/2}\displaystyle%
\prod_{i=1}^{k}(\lambda -\mu _{i}) \\
Q_{k}^{(n)}=\displaystyle\int d\overline{\mu }\Pi _{k}(\overline{\mu }%
),\quad Q_{k}^{(L,n)}=\displaystyle\int_{[-L,L]^{k}}d\overline{\mu }\Pi
_{k}(\overline{\mu }),%
\end{array}%
\end{equation*}%
In particular, $Q_{n}^{(n)}$ is $Q_{n}$ of (\ref{p(la)}). Then, according to
\cite{Sz} (see formulas (2.2.10) - (2.2.11)), we have
\begin{equation}
\psi _{k}^{(n)}(\lambda )=\frac{\gamma _{k}^{(n)}}{Q_{k}^{(n)}}\int \Pi
_{k}^{\ast }(\lambda ,\overline{\mu })\Pi _{k}(\overline{\mu })d\overline{%
\mu },\quad \psi _{k}^{(L,n)}(\lambda )=\frac{\gamma _{k}^{(L,n)}}{%
Q_{k}^{(L,n)}}\int_{[-L,L]^{k}}\Pi _{k}^{\ast }(\lambda ,\overline{\mu })\Pi
_{k}(\overline{\mu })d\overline{\mu },  \label{rep_psi}
\end{equation}%
where $\gamma _{k}^{(n)}$ and $\gamma _{k}^{(L,n)}$ are the coefficients in
front of $\lambda ^{k}$ in $p_{k}^{(n)}(\lambda )$ and $p_{k}^{(L,n)}(%
\lambda )$. They have the form \cite{Sz}
\begin{equation}
(\gamma _{k}^{(n)})^{2}=\frac{Q_{k}^{(n)}(k+1)}{Q_{k+1}^{(n)}},\quad (\gamma
_{k}^{(L,n)})^{2}=\frac{Q_{k}^{(L,n)}(k+1)}{Q_{k+1}^{(L,n)}}.
\label{gamma_k}
\end{equation}%
We prove first that for some $L,\tilde{L}_{1}$ uniformly in $k\leq
(1+\epsilon /4)n$ we have
\begin{equation}
\bigg|\frac{Q_{k}^{(n)}}{Q_{k}^{(L,n)}}-1\bigg|\leq Ce^{-nL_{1}},\quad \bigg|%
\frac{\gamma _{k}^{(n)}}{\gamma _{k}^{(L,n)}}-1\bigg|\leq Ce^{-nL_{1}}.
\label{l*.2}
\end{equation}%
The first relation here follows from the result of \cite{BPS}, Lemma 1,
which states that for any function $V$, satisfying conditions (\ref{log})
with some $\tilde{\epsilon}>0$ and (\ref{Lip}) there exist absolute
constants $L>1$, $\tilde{L}_{1}$ such that for any $k\leq n$
\begin{equation}
\begin{array}{rcl}
\bigg|\displaystyle\frac{\tilde{Q}_{k}^{(n)}}{\tilde{Q}_{k}^{(L,n)}}-1\bigg|
& \leq & Ce^{-n\tilde{L}_{1}}, \\
(1-\chi _{L}(\mu _{1}))\rho _{k}^{(n)}(\mu _{1}) & \leq & C\,\exp \{-n\tilde{%
L}_{1}\log |\mu _{1}|\}, \\
(1-\chi _{L}(\mu _{1}))\rho _{k}^{(n)}(\mu _{1},\mu _{2}) & \leq & C\,\exp
\{-n\tilde{L}_{1}\log |\mu _{1}|\},%
\end{array}
\label{l*.3}
\end{equation}%
where $\chi _{L}$ is the characteristic function of the interval $[-L,L]$
and
\begin{equation}
\rho _{k}^{(n)}(\mu _{1})=(Q_{k}^{(n)})^{-1}\int \Pi _{k}(\overline{\mu }%
)d\mu _{2},\dots d\mu _{k},\quad \rho _{k}^{(n)}(\mu _{1},\mu
_{2})=(Q_{k}^{(n)})^{-1}\int \Pi _{k}(\overline{\mu })d\mu _{3},\dots d\mu
_{k},  \label{def_rho}
\end{equation}%
are the first and the second marginals of the probability density $\Pi _{k}(%
\overline{\mu })/Q_{k}^{(n)}$ of $k$ variables $\overline{\mu }=(\mu
_{1},\dots ,\mu _{k})$, corresponding to the potential $\tilde{V}=nk^{-1}V$
(cf (\ref{p(la)}), (\ref{dp_k})).

Thus, if we chose constants $L,\tilde L_1$ for $\tilde
V(\lambda)=(1+\epsilon/4)V(\lambda)$, $\tilde\epsilon=
\epsilon/4(1+\epsilon/4)$, we obtain the first bound of (\ref{l*.2}). The
second bound follows from the first in view of the relations (\ref{gamma_k}).

Now if we denote by $\Delta _{k}(\lambda )$ the r.h.s. of (\ref{l*.1}),
then, using (\ref{gamma_k}), (\ref{l*.2}), and the second line of (\ref{l*.3}%
), we get
\begin{equation}
\begin{array}{rcl}
\Delta _{k+1}(\lambda ) & \leq & C\,e^{-n\tilde{L}_{1}}+\bigg|\displaystyle%
\frac{\gamma _{k}^{(n)}}{{Q_{k-1}^{(n)}}}\bigg(\int d\overline{\mu }-%
\displaystyle\int_{[-L,L]^{n}}d\overline{\mu }\bigg)\Pi _{k}^{\ast }(\lambda
,\overline{\mu })\,\Pi _{k}(\overline{\mu })\bigg|%
\end{array}
\label{l*.4}
\end{equation}%
Denoting $\Delta _{k}^{\prime }(\lambda )$ the second term in the r.h.s.
of (\ref{l*.4}), we obtain
\begin{equation}
\begin{array}{lll}
\Delta _{k}^{\prime }(\lambda )&\leq& n\displaystyle\frac{\gamma _{k}^{(n)}}{{%
Q_{k}^{(n)}}}\displaystyle\int (1-\chi _{L}(\mu _{1}))\Pi _{k}^{\ast
}(\lambda ,\overline{\mu })\Pi _{k}(\overline{\mu })d\overline{\mu } \\
&\leq& n\sqrt{k+1}\bigg[(Q_{k}^{(n)})^{-1}\int d\overline{\mu }(1-\chi
_{L}(\mu _{1}))\Pi _{k}(\overline{\mu })d\overline{\mu }\bigg]^{1/2} \\
&&\bigg[(Q_{k+1}^{(n)})^{-1}\displaystyle\int d\overline{\mu }(1-\chi _{L}(\mu
_{1}))(\Pi _{k}^{\ast }(\lambda ,\overline{\mu }))^{2}\Pi _{k}(\overline{\mu
})d\overline{\mu }\bigg]^{1/2} \\
&\leq& nL^{1/2}\sqrt{k+1}\bigg|\displaystyle\int (1-\chi _{L}(\mu _{1}))\rho
_{k}^{(n)}(\mu _{1})d\mu _{1}\bigg|^{1/2}\bigg|\displaystyle\int (1-\chi
_{L}(\mu _{1}))\rho _{k+1}^{(n)}(\lambda ,\mu _{1})d\mu _{1}\bigg|^{1/2}%
\end{array}
\label{l*.5}
\end{equation}%
Here we have used (\ref{gamma_k}) and (\ref{def_rho}). According to (\ref%
{l*.3}) the integrals in the r.h.s. of (\ref{l*.5}) are $O(n^{-1}e^{-n\tilde{%
L}_{1}|\log L|})$. Thus, taking $L_{1}=\tilde{L}_{1}\log L/2$, we get (\ref%
{l*.1}) if $n$ is large enough.

It follows from (\ref{rec}) that
\begin{equation*}
J_{k}^{(n)}=\gamma _{k}^{(n)}/\gamma _{k+1}^{(n)}.
\end{equation*}%
This and (\ref{l*.2}) imply (\ref{JL}). Another way to prove (\ref{JL}) is
to apply (\ref{l*.1}) and the second line of (\ref{l*.3}) to formula (\ref%
{Jpp}).

The proof of (\ref{KL}) follows from (\ref{k_n}) and (\ref{l*.1})

\medskip

\noindent \textit{Proof of Proposition \ref{pro:d1}}. By general principles
(see e.g.\cite{At}) the kernel $R_{\zeta }(x,y)$ of the resolvent of
differential operator (\ref{Aop}) has the form
\begin{equation*}
R_{\zeta }(x,y)=\frac{1}{W}\left\{
\begin{array}{cc}
\psi_{-}(x)\psi_{+}(y) & x\leq y, \\
\psi_{-}(y)\psi_{+}(x) & x\geq y,%
\end{array}%
\right.
\end{equation*}%
where $f_{\pm }(x,\zeta )$ are solutions of the differential equations
\begin{equation*}
\frac{a}{2}\psi^{\prime \prime }(x)-(2cx+\zeta )\psi(x)=0,
\end{equation*}%
that are square integrable at $\pm \infty $, and $W$ is fixed be the
condition
\begin{equation*}
\frac{\partial }{\partial x}R_{z}(x+0,x)-\frac{\partial }{\partial x}%
R_{z}(x-0,x)=\frac{2}{a}.
\end{equation*}%
According to formulas (\ref{as_AC}), we can choose $\psi_{+}(x)=\,\hbox{Ai}%
(\kappa x+\gamma \zeta )$ and $\psi_{-}(x)=\hbox{Ci}(\kappa x+\gamma \zeta )=i%
\hbox{Ai}(\kappa x+\gamma \zeta )+\hbox{Bi}(\kappa x+\gamma \zeta )$. This
and the identity (see \cite{AS}),
\begin{equation*}
(\hbox{Ai})^{\prime }(z)\hbox{Ci}(z)-\hbox{Ai}(z)(\hbox{Ci})^{\prime
}(z)=\pi ^{-1}.
\end{equation*}%
lead to (\ref{R-psi}) - (\ref{Ci}).

Relations (\ref{as_psi+}), (\ref{as_psi-}) follow from the asymptotic
representations (see \cite{AS})
\begin{equation}
\begin{array}{rcll}
\hbox{Ai}(z) & = & \pi ^{-1/2}z^{-1/4}e^{-\frac{2}{3}z^{3/2}}\bigg(%
1+O(z^{-3/2})\bigg), & |\hbox{arg}z|<\pi , \\
\hbox{Ai}(-z) & = & \pi ^{-1/2}z^{-1/4}\sin \bigg(\displaystyle\frac{2}{3}%
z^{3/2}+\frac{\pi }{4}\bigg)\bigg(1+O(z^{-3/2})\bigg), & |\hbox{arg}z|<%
\displaystyle\frac{2}{3}\pi , \\
\hbox{Ci}(z) & = & \pi ^{-1/2}z^{-1/4}e^{\frac{2}{3}z^{3/2}}\bigg(%
1+O(z^{-3/2})\bigg), & |\hbox {arg}z|<\displaystyle\frac{\pi }{3}, \\
\hbox{Ci}(-z) & = & \pi ^{-1/2}z^{-1/4}\exp \bigg\{i\bigg(\displaystyle\frac{%
2}{3}z^{3/2}+\frac{\pi }{4}\bigg)\bigg\}\bigg(1+O(z^{-3/2})\bigg), & |%
\hbox{arg}z|<\displaystyle\frac{2}{3}\pi .%
\end{array}
\label{as_AC}
\end{equation}%
The leading terms of the asymptotic formulas for the derivatives of $\hbox{Ai}$ and $\hbox{Ci}$ can
be obtained as the leading terms of the  formal derivatives at the above formulas. This and (\ref%
{R-psi}) lead to  assertion $(iii)$ of the proposition.

Assertion (iv) follows from (ii) - (iii) and (\ref{as_AC}).

\medskip

\noindent \textit{Proof of Lemma \ref{lem:1}} It is easy to see, that the
l.h.s. of (\ref{est2}) is bounded above by the expression
\begin{equation}
\displaystyle\sum_{k=0}^{n-M}{\sum_{m=0}^{\infty
}|R_{k,m}^{(n)}(z)|^{2}=\sum_{k=1}^{n-M}(R^{(n)}(z)R^{(n)}(}\overline{z}{))}%
_{kk}={(n-M)\int \frac{\rho _{n-M}^{(n)}(\lambda)d\lambda}{|x-z|^{2}},}  \label{l1.1}
\end{equation}%
where $\rho _{n-M}^{(n)}(\lambda)$ is the first marginal density of the
probability density (cf (\ref{p(la)}))
\begin{equation}
p_{n-M}^{(n)}(\lambda _{1},...\lambda _{n-M})=\left( Q_{M-n}^{(n)}\right)
^{-1}\prod_{1\leq j<k\leq n-M}(\lambda _{j}-\lambda _{k})^{2}\exp
\{-(n-M)\sum_{j=1}^{n-M}\frac{1}{1-M/n}\cdot V(\lambda _{j})\}  \label{pMn}
\end{equation}%
This suggest the introduction of the functional
\begin{equation}
\mathcal{E}_{\delta }[m]=\frac{1}{1-\delta }\int V(\lambda )m(d\lambda
)-\int \int \log |\lambda -\mu |m(d\lambda )m(d\mu )  \label{End}
\end{equation}%
with $\delta \in (0,1)$ (the functional (\ref{En})) corresponds to the case $%
\delta =0$ of the above functional. According to the results of \cite{BPS}
(see also \cite{Jo:98}), if $V$ satisfies (\ref{log}), and $V^{\prime }$ is
a locally H\"{o}lder function, then  the unique minimizer of the functional
(\ref{End}) is equal to the limit $\rho ^{(\delta )}$ of the first marginal
density $\rho _{n-M}^{(n)}$ of the distribution (\ref{pMn}) as $n\rightarrow
\infty ,\;M\rightarrow \infty ,\;M/n\rightarrow \delta $, the limit is in a
certain \textquotedblright energy\textquotedblright\ norm, determined by the
second term of (\ref{End}). It is easy to find (see also \cite%
{Sa-To:97,Kr-Mc:00}), that the support $\sigma _{\delta }$ of the density $%
\rho ^{(\delta )}$ lies strictly inside of the support $\sigma $ of the DOS
(i.e., the density of the limit of the first marginal density of the
distribution (\ref{p(la)}), corresponding to $\delta =0$ in (\ref{End})).
Moreover, the endpoint $a_{\delta }$ of $\sigma _{\delta }$ and the endpoint
$a$ of $\sigma $ are related as follows
\begin{equation}
\lim_{\delta \rightarrow 0}(a-a_{\delta })\delta ^{-1}=K>0,  \label{aad}
\end{equation}%
where $K$ can be written via derivatives of the function $u(x)$, defined in (%
\ref{u}) (this relation can also be deduced from (\ref{quaeq'}) in \cite{APS}).

Now we need the following proposition, proven in \cite{APS} (see the last
inequality in the proof of Proposition 2 and the text before the inequality).

\noindent

\begin{proposition}
\label{pro:1} Consider a unitary invariant ensemble of the form (\ref{p(M)}%
)--(\ref{dM}) and assume that $V(\lambda )$ possesses two bounded derivatives
in some neighborhood of the support $\sigma $ of the Density of States $\rho
$, that satisfies condition C2. Denote by $\sigma (\varepsilon )$ the $%
\varepsilon $-neighborhood of the spectrum $\sigma $. Then there exist $%
C_{1} $, $C_{2}$
\begin{equation}
\int_{\mathbf{R}\setminus \sigma (C_{1}n^{-1/2}\log n)}\rho _{n}(\lambda
)d\lambda \leq e^{-C_{2}\,n^{1/2}\log n} . \label{est1}
\end{equation}
\end{proposition}

According to Proposition \ref{pro:1}, we have uniformly in $0\leq M/n\leq
\delta _{0}<1$
\begin{equation}
\int_{a_{\delta }+C_{1}n^{-1/2}\log n}\rho _{n-M}^{(n)}(\lambda )d\lambda
\leq \exp \{-C_{2}n^{1/2}\log n\},  \label{l1.2}
\end{equation}%
where $C_{1}$ and $C_{2}$ may depend on $\delta _{0}$.

Set $\varepsilon (n)=KM/2n$. Then $\varepsilon (n)>>C_{1}n^{-1/2}\log n$,
and so $a-\varepsilon (n)>a_{\delta }+C_{1}n^{-1/2}\log n$ for sufficiently
big $n$ and $M>>n^{1/2}\log n$. Thus, we can write:
\begin{equation}
\begin{array}{l}
\displaystyle{\int {\frac{\rho _{n-M}^{(n)}(\lambda )d\lambda }{%
|a+n^{-2/3}\zeta -\lambda |^{2}}}=\bigg[\int_{\lambda >a-\varepsilon
(n)}+\int_{\lambda \leq a-\varepsilon (n)}\bigg]{\frac{\rho
_{n-M}^{(n)}(\lambda )d\lambda }{(a+n^{-2/3}\zeta -\lambda )^{2}}}} \\
\displaystyle{\leq {\frac{n^{4/3}}{|\Im \zeta |^{2}}}\exp
\{-C_{0}n^{1/2}\log n\}+\int_{\lambda \leq a-\varepsilon (n)}\frac{\rho
_{n-M}^{(n)}(\lambda )d\lambda }{|\lambda -a-n^{-2/3}\Re \zeta
|^{2}+n^{-4/3}|\Im \zeta |^{2}}} \\
\displaystyle{\leq O(e^{-C\cdot \sqrt{n}})+C\int \frac{\rho
_{n-M}^{(n)}(\lambda )d\lambda }{|\lambda -a|^{2}+(M/n)^{2}}\leq
O(e^{-C\cdot \sqrt{n}})+C|\Im g_{M-n}^{(n)}(a+iM/n)|n/M,}%
\end{array}
\label{l1.3}
\end{equation}%
where $g_{n-M}^{(n)}(z)$ is the Stieltjes transform of the measure $\rho
_{n-M}^{(n)}(\lambda )d\lambda $ (see (\ref{g(z)})). According to the
results of \cite{PS:97}, Eqs.(2.5) - (2.6), if $z\in \mathbf{D}$, then $%
g_{n-M}^{(n)}(z)$ verifies the relation
\begin{equation}
(g_{n-M}^{(n)}(z))^{2}-\frac{1}{1-M/n}V^{\prime }(z)g_{n-M}^{(n)}(z)+%
\mathcal{Q}_{n}^{(M/n)}(z)=O(\frac{1}{n^{2}|\Im z|^{4}}),\;n\rightarrow
\infty ,  \label{quaeq}
\end{equation}%
where (cf (\ref{Q}))
\begin{equation}
\mathcal{Q}_{n}^{(M/n)}(z)=\frac{1}{1-M/n}\int \frac{V^{\prime
}(z)-V^{\prime }(x)}{z-\lambda }\rho _{n-M}^{(n)}(\lambda )d\lambda .
\label{QMn}
\end{equation}%
On the other hand, if we consider $g^{(\delta )}(z)$, the Stieltjes
transform of the measure $\rho ^{(\delta )}(\lambda )d\lambda $, minimizing (%
\ref{End}), then for $z\in \mathbf{D}$ $g^{(\delta )}(z)$ is a solution of
the quadratic equation (see (\ref{g,P})--(\ref{Q})%
\begin{equation}
(g^{(\delta )}(z))^{2}-\frac{1}{1-\delta }\mathcal{V}^{\prime }(z)g^{(\delta
)}(z)+\mathcal{Q}^{(\delta )}(z)=0  \label{quaeq'}
\end{equation}%
in the class of analytic functions, such that $\Im z\cdot \Im g>0$, and $%
\mathcal{Q}^{(\delta )}(z)$ is defined by the formula (\ref{QMn}) with $\rho
^{(\delta )}$ instead of $\rho _{n-M}^{(n)}$ and $\delta $ instead of $M/n.$
Now (\ref{quaeq})--(\ref{quaeq'}) and the fact that the analytic function $(%
\mathcal{V}^{\prime }(z)(1-\delta )^{-1})^{2}-4\mathcal{Q}^{(\delta )}(z)$
is zero at $z=a_{\delta }$ imply
\begin{equation*}
|\Im g_{n-M}^{(n)}(z)|\leq C(|\Im z|+|z-a_{\delta }|^{1/2}+|\mathcal{Q}%
_{n}^{(M/n)}(z)-\mathcal{Q}^{(M/n)}(z)|^{1/2}+n^{-1}|\Im z|^{-2}).
\end{equation*}%
Besides, since $(\mathcal{V}^{\prime }(z)-V^{\prime }(\lambda ))/(z-\lambda
) $ has bounded derivative with respect to $\lambda $ (recall that $\mathcal{%
V} $ is analytic in a certain neighborhood of the support ) we can use
results of paper \cite{BPS}, according to which
\begin{equation*}
|\mathcal{Q}_{n}^{(M/n)}(z)-\mathcal{Q}^{(M/n)}(z)|\leq \,C\cdot
n^{-1/2}\log ^{1/2}n.
\end{equation*}%
The last two inequalities and (\ref{aad}) yield%
\begin{equation}\label{l1.4}
|\Im g_{n-M}^{(n)}(a+iM/n)|\leq C\cdot ((M/n)^{1/2}+n^{-1/4}\log
^{1/4}n+n/M^{2}).
\end{equation}
Now, using (\ref{l1.3}) and (\ref{l1.4}), we obtain from (\ref{l1.3})
\begin{equation*}
\int {\frac{\rho _{n-M}^{(n)}(\lambda )d\lambda }{|a_{j}+n^{-2/3}\zeta
-\lambda |^{2}}}\leq \,C\cdot ((n/M)^{1/2}+n^{2}/M^{3}+n^{3/4}\log
^{1/4}n/M).
\end{equation*}%
Combining this bound with (\ref{l1.1}), we obtain (\ref{est2}). Lemma \ref%
{lem:1} is proved.

\bigskip

\textit{Proof of Lemma \ref{lem:2}}
Take $\nu \in \mathbf{N}$ to provide the bound (see Lemma \ref{lem:D})
\begin{equation}
||D^{(M)}||^{\nu }\leq Cn^{-2}.  \label{nu}
\end{equation}%
By (\ref{RRD}) we have
\begin{equation}
R^{(n)}=R^{\ast }+\sum_{l=1}^{\nu -1}(-1)^{l}R^{\ast }D^{l}+(-1)^{\nu
}R^{(n)}D^{\nu },  \label{RRD-n}
\end{equation}%
We write for both $R=R^{(n)}$ and $R=R^{\ast }$ and $l=1,...,p$:
\begin{equation}
(RD^{l})_{n-j,n-k}=\displaystyle\sum_{m_{1},\dots ,m_{\ell
}=-2M}^{2M}R_{n-j,n-m_{1}}D_{n-m_{1},n-m_{1}}\dots
D_{n-m_{l},n-k}+d_{n-j,n-k}^{(l)}  \label{l2.1}
\end{equation}%
with
\begin{equation*}
\begin{array}{lll}
|d_{n-j,n-k}^{(l)}| & = & \left\vert \sum^{\prime
}R_{n-j,n-m_{1}}D_{n-m_{1},n-m_{1}}\dots D_{n-m_{l},n-k}\right\vert \\
&  & \leq \displaystyle\sum_{p=1}^{l}\displaystyle%
\sum_{|m|>2M}(|R||D|^{l-p})_{n-j,n-m}(|D|)_{n-m,n-k}^{p},%
\end{array}%
\end{equation*}%
where $\sum^{\prime }$ is the sum of the terms which contain at least one $%
|m_{i}|>2M$ and $|R|_{l,l^{\prime }}=|R_{l,l^{\prime }}|$, $|D|_{l,l^{\prime
}}=|D_{l,l^{\prime }}|$.

We observe that for $m\neq m^{\prime }$ we have
\begin{subequations}
\begin{equation}
|D|_{n-m,n-m^{\prime }}\leq \max_{i}J_{i}^{(n)}(|R_{n-m-1,n-m^{\prime
}}^{\ast }(z)|+|R_{n-m+1,n-m^{\prime }}^{\ast }(z)|)+2a|R_{n-m,n-m^{\prime
}}^{\ast }(z)|),  \label{b_D}
\end{equation}%
where we also took into account that since the matrix $J^{(n)}$ corresponds
now to the polynomials $p^{(L,n)}$, orthogonal on a finite interval $[-L,L]$
(see Lemma \ref{lem:*}), the coefficients $J_{i}(n)$ are bounded uniformly
in $i$ and $n$. Hence, according to (\ref{as_psi+}) and (\ref{as_psi-}), for
any $p=1,\dots ,\nu $, $|k|\leq m$, $|m|>2M$ we obtain for some positive $%
\alpha $ and $\beta $
\end{subequations}
\begin{equation}
\begin{array}{lll}
(|D|^{p})_{n-m,n-k} & \leq & C\,n^{\alpha }\exp \{-C\varepsilon
(Mn^{-1/3})^{1/2}-C\varepsilon ((|m|-2M)n^{-1/3})^{1/2}\} \\
& \leq & C\,e^{-C\varepsilon n^{2/15}}\exp \{-C\varepsilon
((|m|-2M)n^{-1/3})^{1/2}\},%
\end{array}
\label{b_D^l}
\end{equation}%
and
\begin{equation}
|(|D|^{p}(|D|^{\dag })^{p})_{n-m,n-m}|\leq C\,n^{\alpha }m^{\beta },\;|m|>2M.
\label{b2_D^l}
\end{equation}%
We will also use the trivial bounds: $||R^{(n)}||\leq n^{2/3}\varepsilon
^{-1}$,
\begin{equation*}
\begin{array}{lll}
|(|R^{(n)}||D|^{p})_{n-j,n-m}| & \leq & ||R^{(n)}||\cdot
|(|D|^{p}(|D|^{\dag })^{p})_{n-m,n-m}|^{1/2}, \\
&  &  \\
|(R^{\ast }|D|^{p})_{n-j,n-m}| & \leq & (|R^{\ast }||R^{\ast }|^{^{\dag
}})_{n-j,n-j}^{1/2}\cdot |(|D|^{p}(|D|^{\dag })^{p})_{n-m,n-m}|^{1/2}. \\
&  &
\end{array}%
\end{equation*}%
The above bounds yield
\begin{equation*}
|d_{n-j,n-k}^{(l)}|\leq Ce^{-C\varepsilon (Mn^{-1/3})^{1/2}},
\end{equation*}%
and we have from (\ref{RRD-n}) and (\ref{l2.1})%
\begin{equation}
\begin{array}{lll}
R_{n-j,n-k}^{(n)} & = & R_{n-j,n-k}^{\ast }+\displaystyle\sum_{l =1}^{\nu
-1}(-1)^{l}(R^{\ast }(D^{(M)})^{l})_{n-j,n-k} \\
&  & +(-1)^{\nu }(R^{(n)}(D^{(M)})^{\nu })_{n-j,n-k}+O(e^{-C\varepsilon
(Mn^{-1/3})^{1/2}}).%
\end{array}
\label{12.2}
\end{equation}%
Thus, we get for sufficiently large $n$
\begin{equation}
\begin{array}{c}
n^{-4/3}\displaystyle\sum_{j=1}^{M}\displaystyle\sum_{|k|\leq
M}|R_{n-j,n-k}^{(n)}(z)|^{2}=n^{-4/3}\displaystyle\sum_{j=1}^{M}%
\displaystyle\sum_{|k|\leq 2M}|R_{n-j,n-k}^{\ast }(z)|^{2}+\delta _{n}(z),%
\end{array}%
\end{equation}%
where
\begin{equation*}
\begin{array}{lll}
|\delta _{n}(z)| & \leq & Cn^{-4/3}\sum_{\ell =1}^{\nu -1}\displaystyle%
\sum_{j=1}^{M}(R^{\ast }(D^{(M)})^{l}(D^{(M)\dag })^{l}R^{\ast \dag
})_{n-j,n-j} \\
&  & +C\displaystyle\sum_{j=1}^{M}(R^{(n)}(D^{(M)})^{\nu }(D^{(M)\dag
})^{\nu }R^{(n)\dag })_{n-j,n-j}+o(n^{-1}) \\
& \leq & 2C||D^{(M)}||^{2}n^{-4/3}\displaystyle\sum_{j=1}^{M}\displaystyle%
\sum_{|k|\leq 2M}|R_{n-j,n-k}^{\ast }|^{2}+C||D^{(M)}||^{2\nu }\displaystyle%
\sum_{j=1}^{M}(R^{(n)}R^{(n)\dag })_{n-j,n-j}+o(n^{-1}) \\
& \leq & 2C||D^{(M)}||^{2}n^{-4/3}\displaystyle\sum_{j=1}^{M}\displaystyle%
\sum_{|k|\leq 2M}|R_{n-j,n-k}^{\ast }|^{2}+C\varepsilon
^{-2}||D^{(M)}||^{2\nu }Mn^{4/3}+o(n^{-1})%
\end{array}%
\end{equation*}%
Then, in view of (\ref{nu}), we obtain (\ref{est3}). Inequality (\ref{est3a})
can be proved similarly.\bigskip

\noindent \textit{Proof of Lemma \ref{lem:D}.} By the direct calculation
we find from (\ref{Ji}) for $j\not=k$:
\begin{equation}
\begin{array}{c}
((J^{(n)}-zI)R^{\ast }(z))_{n-j,n-k}=-aR_{n-j,n-k}^{\ast }(z)+\fracd{a}{2}%
R_{n-j+1,n-k}^{\ast }(z)+{\fracd{a}{2}}R_{n-j-1,n-k}^{\ast }(z) \\
\displaystyle{-n^{-2/3}\zeta R_{n-j,n-k}^{\ast }(z)-c{\frac{j}{n}}%
R_{n-j+1,n-k}^{\ast }(z)-c{\frac{j-1}{n}}R_{n-j-1,n-k}^{\ast }(z)}\nonumber
\\
\displaystyle{+r_{j}^{(n)}R_{n-j+1,n-k}^{\ast
}(z)+r_{j-1}^{(n)}R_{n-j-1,n-k}^{\ast }(z)+r_{j}^{(n)}R_{n-j,n-k}^{\ast }(z).%
}\nonumber%
\end{array}
\label{JR}
\end{equation}%
By using the Taylor formula of the forth order for the second and the third
terms and the same formula of the second order for the fifth and the sixth
terms, we rewrite the r.h.s of the last formula as
\begin{equation}
n^{-1/3}\bigg(-\zeta R_{\zeta }(x,y)+{\frac{a}{2}}{\frac{\partial ^{2}}{%
\partial x^{2}}}R_{\zeta }(x,y)-2cxR_{\zeta }(x,y)\bigg)\bigg|_{x={\frac{j}{%
n^{1/3}}},y={\frac{k}{n^{1/3}}}}+{d}_{n-j,n-k}(z).  \label{JR1}
\end{equation}%
The expression in the brackets of the last equality is equal to zero because
of equation (\ref{eqRz}) for $x\neq y$. The remainder ${d}_{n-j,n-k}(z)$ is
a linear combination of
\begin{equation*}
 {{\frac{\partial ^{\alpha }}{\partial x^{\alpha }}}R_{\zeta
}(x,k/n^{1/3})},\quad\alpha =0,2,4,
\end{equation*}%
with $x=(j+\theta )/n^{2/3}$, where $|\theta |\leq 1$ can be different for
different terms. The derivatives can be excluded by using equation (\ref%
{eqRz}) with corresponding $\theta $. This leads to the bound for the
remainder ${{d}_{n-j,n-k}(z)}$ in (\ref{JR1}):
\begin{equation}\begin{array}{lll}
|{d}_{n-j,n-k}(z)|&\leq& C\,\bigg\{\fracd{(j/n^{1/3})^{2}+|\zeta |^{2}+1}{n}
\max_{|\theta |<1}|{R_{\zeta }}((j+\theta)/n^{1/3},
k/n^{1/3})|\\
&&\ds{+n^{-1}|j/n^{1/3}|\max_{|\theta |<1}\bigg|{{\fracd{\partial }{\partial x}}R_{\zeta }}%
((j+\theta)/n^{1/3},k/n^{1/3})\bigg|\bigg\}.}
  \label{ti_d}
\end{array}\end{equation}%
Similarly
\begin{equation}
\begin{array}{l}
((J^{(n)}-zI)R^{\ast })_{n-k,n-k}=-aR_{n-k,n-k}^{\ast }(z)+{\displaystyle%
\frac{a}{2}}R_{n-k+1,n-k}^{\ast }(z)+{\displaystyle\frac{a}{2}}%
R_{n-k-1,n-k}^{\ast }(z) \\
-n^{-2/3}\zeta R_{n-k,n-k}^{\ast }(z)-c{\displaystyle\frac{k}{n}}%
R_{n-k+1,n-k}^{\ast }(z)-c{\displaystyle\frac{k-1}{n}}R_{n-k-1,n-k}^{\ast
}(z) \\
+r_{k}^{(n)}R_{n-k+1,n-k}^{\ast }(z)-r_{k-1}^{(n)}R_{n-k-1,n-k}^{\ast }(z)+%
\overline{r}_{k}^{(n)}R_{n-k,n-k}^{\ast }(z) \\
=\bigg(-{\displaystyle\frac{a}{2}}{\displaystyle\frac{\partial }{\partial x}}%
R_{z}(x-0,y)+{\displaystyle\frac{a}{2}}{\displaystyle\frac{\partial }{%
\partial x}}R_{z}(x+0,y)\bigg)\bigg|_{x=y=kn^{-1/3}}+{d}_{n-k,n-k}(z).%
\end{array}
\label{JRd}
\end{equation}%
The expression in the square brackets is $1$ because of equation (\ref{eqRz}%
), and the remainders ${d}_{k,k}(z)$ admits the bound
\begin{equation}
|{d}_{n-k,n-k}(\zeta )|\leq C\,n^{-1/3}\bigg\{(|k/n^{1/3}|+|\zeta |+1)\max_{|\theta
|\leq 1}|{R_{\zeta }}((k+\theta )/n^{1/3},k/n^{1/3})|\bigg\}.
\label{ti_dd}
\end{equation}%
 We will use the bound
\begin{equation}
||\tilde{D}(z)||\leq ||\tilde{D}(z)||_{1}=\left( \max_{j}\sum_{k}|{d}%
_{n-j,n-k}|\max_{k}\sum_{j}|{d}_{n-j,n-k}|\right) ^{1/2}.  \label{D.0}
\end{equation}%
First, by using Proposition \ref{pro:d1}, it is easy to show that
\begin{equation*}
\max_{|k|\leq 2M}|{d}_{n-k,n-k}|\leq C\,\,\cdot n^{-1/3}\max_{|x|\leq \mu
}|x||\psi _{+}(x,\zeta )||\psi _{-}(x,\zeta )|,
\end{equation*}%
where
\begin{equation}
\mu =Mn^{-1/3}.  \label{mu}
\end{equation}%
By using asymptotics (\ref{as_psi+}) and (\ref{as_psi-}), we find that
\begin{equation}
\max_{|k|\leq 2M}|{d}_{n-k,n-k}|=O(\mu ^{1/2}n^{-1/3})=O((M/n)^{1/2}).
\label{D.1}
\end{equation}%
Now we have to estimate $\max_{|j|<M}\displaystyle\sum_{k\not=j}|\tilde{d}%
_{n-jn-,k}|$, and $\max_{|k|<2M}\displaystyle\sum_{j\not=k}|\tilde{d}%
_{n-j,n-k}|$. A standard, but tedious analysis, based on (\ref{ti_d}) and
Proposition \ref{pro:d1}, shows that the leading contribution to these
quantities is due to the expression
\begin{equation*}
n^{-1}\max_{|j|<2M}(jn^{-1/3})^{2}|\psi _{-}(jn^{-1/3},\zeta
)|\sum_{k=j}^{2M}|\psi _{+}(kn^{-1/3},\zeta )|
\end{equation*}%
which is asymptotically equivalent to
\begin{equation*}
n^{-2/3}\max_{|x|\leq \mu }x^{2}|\psi _{-}(x,\zeta )|\int_{x}^{\mu }|\psi
_{+}(y,\zeta )|dy\leq C\,n^{-2/3}\mu ^{2}\varepsilon ^{-1}=C\,\varepsilon
^{-1}n^{-4/3}M^{2}.
\end{equation*}%
Lemma is proved.

\medskip

\medskip
\noindent \textit{Proof of Lemma \ref{pro:3}.} Let us chose
\begin{equation}
M_{1}=[n^{1/2}\log ^{6}n],\,\,\varepsilon =n^{-1/12}\log ^{-1}n,\,\,L=\bigg[%
C_{1}\frac{n^{-1/2}\log n}{n^{-2/3}\varepsilon }\bigg],  \label{p3.1}
\end{equation}%
where $C_{1}$ is defined in Proposition \ref{pro:1}. Then, by Proposition %
\ref{pro:1}, we have
\begin{equation*}
n\int_{a+n^{-2/3}L_{0}}^{\infty }\rho _{n}(\lambda )d\lambda
=n\int_{a+n^{-2/3}L_{0}}^{a+n^{-2/3}\varepsilon L}\rho _{n}(\lambda
)d\lambda +o(1).
\end{equation*}%
Besides, similarly to the proof of Lemma \ref{lem:1}, if we consider
\begin{equation*}
\rho _{n-M_{1}}^{(n)}(\lambda )=(n-M_{1})^{-1}\sum_{j=1}^{n-M_{1}}\psi
_{j}^{2}(\lambda ),
\end{equation*}%
(we omitted the superscript ${(n)}$ in $\psi $'s) then we obtain in view of
the same proposition
\begin{equation*}
(n-M_{1})\int_{a+n^{-2/3}L_{0}}^{a+n^{-2/3}\varepsilon L}\rho
_{n-M_{1}}^{(n)}(\lambda )d\lambda \leq e^{-C\sqrt{n}}.
\end{equation*}%
So
\begin{equation}
\begin{array}{c}
\displaystyle{\ n\int_{a+n^{-2/3}L_{0}}^{\infty }\rho _{n}(\lambda )d\lambda
=\int_{a+n^{-2/3}L_{0}}^{a+n^{-2/3}\varepsilon L}\sum_{k=1}^{M_{1}}\psi
_{n-k}^{2}(\lambda )d\lambda +o(1)} \\
\displaystyle{\ =\sum_{p=[L_{0}/\varepsilon
]}^{L-1}\int_{a+n^{-2/3}\varepsilon p}^{a+n^{-2/3}\varepsilon
(p+1)}\sum_{k=1}^{M_{1}}\psi _{n-k}^{2}(\lambda )d\lambda
+o(1)=\sum_{p=[L_{0}/\varepsilon ]}^{L-1}I_{p}+o(1)}.%
\end{array}
\label{p3.2}
\end{equation}%
The term $I_{p}$ of the sum in the r.h.s. admits the bound:
\begin{equation}
\begin{array}{lcl}
I_{p} & \leq & \displaystyle{\ 2\varepsilon
^{2}n^{-4/3}\sum_{k=1}^{M_{1}}\int_{a+n^{-2/3}\varepsilon
p}^{a+n^{-2/3}\varepsilon (p+1)}\frac{\psi _{n-k}^{2}(\lambda )d\lambda }{%
|\lambda -a-p\varepsilon n^{-2/3}|^{2}+n^{-4/3}\varepsilon ^{2}}} \\
& \leq & \displaystyle{2\varepsilon n^{-2/3}\sum_{k=1}^{M_{1}}|\Im
R_{n-k,n-k}^{(n)}(a+n^{-2/3}\zeta _{p})|} \\
& \leq & \displaystyle{2\varepsilon n^{-2/3}\sum_{k=1}^{M_{1}}|\Im
R_{n-k,n-k}^{\ast }(a+n^{-2/3}\zeta _{p})|} \\
&  & +2\displaystyle{\varepsilon
n^{-2/3}\sum_{k=1}^{M_{1}}%
\sum_{j=-2M_{1}}^{2M_{1}}|R_{n-k,n-j}^{(n)}(a+n^{-2/3}\zeta _{p})\tilde{D}%
_{n-j,n-k}(a+n^{-2/3}\zeta _{p})|} \\
&  & \displaystyle{+2\varepsilon
n^{-2/3}\sum_{k=1}^{M_{1}}\sum_{|j|>2M_{1}}|R_{n-k,n-j}^{(n)}(a+n^{-2/3}%
\zeta _{p})D_{n-j,n-k}(a+n^{-2/3}\zeta _{p})|} \\
& = & \Sigma _{p1}+\Sigma _{p2}+\Sigma _{p3},%
\end{array}
\label{p3.3}
\end{equation}%
where we denote $\zeta_p=\varepsilon(p+i)$ and $D^{(M_1)}$ is defined by (%
\ref{ti-D}) with $M_1$ instead of $M$ of (\ref{M}).

By (\ref{R}) and (\ref{R-psi}),
\begin{equation}
\begin{array}{lcl}
\Sigma _{p1} & = & \displaystyle\frac{4\pi }{\kappa a}\varepsilon n^{-1/3}%
\displaystyle\sum_{k=1}^{M_{1}}|\Im \hbox{Ai}(\kappa \frac{k}{n^{1/3}}%
+\gamma \zeta _{p})\hbox{Ci}(\kappa \frac{k}{n^{1/3}}+\gamma \zeta _{p})| \\
& \leq & \displaystyle\frac{4\pi }{\kappa a}\varepsilon n^{-1/3}\displaystyle%
\sum_{k=1}^{M_{1}}\left( |\Re \hbox{Ai}^{2}(\kappa \frac{k}{n^{1/3}}+\gamma
\zeta _{p})|+|\Im \hbox{Ai}(\kappa \frac{k}{n^{1/3}}+\gamma \zeta _{p})%
\hbox{Bi}(\kappa \frac{k}{n^{1/3}}+\gamma \zeta _{p})|\right) .%
\end{array}
\label{p3.5}
\end{equation}%
Furthermore, by the Schwartz inequality \ref{lem:2}, we get
\begin{equation}
\begin{array}{lcl}
\Sigma _{p2} & \leq & C\,\varepsilon n^{-2/3}\displaystyle\sum_{k=1}^{M_{1}}%
\displaystyle\sum_{j=-2M_{1}}^{2M_{1}}|R_{n-k,n-j}^{(n)}D_{n-j,n-k}| \\
& \leq & C\,\varepsilon \bigg[n^{-4/3}\displaystyle\sum_{k=1}^{2M_{1}}%
\displaystyle\sum_{j=-2M_{1}}^{2M_{1}}|R_{n-j,n-k}^{(n)}|^{2}\bigg]^{1/2}%
\bigg[\displaystyle\sum_{k=1}^{M_{1}}\displaystyle%
\sum_{j=-2M_{1}}^{2M_{1}}|D_{n-j,n-k}|^{2}\bigg]^{1/2}.%
\end{array}
\label{p3.6}
\end{equation}%
For the first factor we use Lemma \ref{lem:2} in which $M$ of (\ref{M}) is
replaced by $2M_{1}$ of (\ref{p3.1}) and $\varepsilon $ is also given by (%
\ref{p3.1}). Since in this case we have by (\ref{||D||}) $%
||D^{(2M_{1})}||=O(n^{-1/4}\log ^{13}n)$, the conditions of the lemma are
satisfied and we obtain in view of (\ref{R})
\begin{equation*}
\begin{array}{lcl}
n^{-4/3}\displaystyle\sum_{k=1}^{2M_{1}}\displaystyle\sum_{|j|\leq
2M_{1}}|R_{n-k,n-j}^{(n)}|^{2} & \leq & 3n^{-2/3}\displaystyle%
\sum_{k=1}^{2M_{1}}\sum_{|j|\le 4M_{1}}\left\vert R_{\zeta _{p}}\left( \frac{k}{%
n^{1/3}},\frac{j}{n^{1/3}}\right) \right\vert ^{2}+o(n^{-1}). \\
&  &
\end{array}%
\end{equation*}%
Besides, we have in view of (\ref{ti_d}):
\begin{eqnarray*}
\sum_{k=1}^{M_{1}}\sum_{j=-2M_{1}}^{2M_{1}}|D_{n-k,n-j}|^{2} &\leq &C\frac{%
M_{1}^{4}}{n^{10/3}}\displaystyle\sum_{k=1}^{M_{1}}\sum_{j=-2M_{1}}^{2M_{1}}%
\left\vert R_{\zeta _{p}}\left( \frac{k}{n^{1/3}},\frac{j}{n^{1/3}}\right)
\right\vert ^{2}+C\frac{M_{1}^{2}}{n^{4/3}}\,\displaystyle%
\sum_{k=1}^{M_{1}}\left\vert R_{\zeta _{p}}\left( \frac{k}{n^{1/3}},\frac{j}{%
n^{1/3}}\right) \right\vert ^{2} \\
&&+C\frac{M_{1}^{2}}{n^{8/3}}\displaystyle\sum_{k=1}^{M_{1}}%
\sum_{j=-2M_{1}}^{2M_{1}}\left\vert  \frac{\partial }{\partial x}%
R_{\zeta _{p}}\left(\frac{k}{n^{1/3}} ,\frac{j}{n^{1/3}}\right) \right\vert ^{2}.
\end{eqnarray*}%

Furthermore, by using representations (\ref{R}) and
(\ref{as_psi-}), we obtain for any $0\leq k\leq M_{1}$,
$|j|>2M_{1}$ and sufficiently large $n$
\begin{equation}
|D_{n-j,n-k}^{(M_{1})}(a+n^{-2/3}\zeta _{p})|\leq n^{1/3}e^{-C\varepsilon
(M_{1}n^{-1/3})^{1/2}}\exp \bigg\{-C\frac{||j|-2M_{1}|^{1/2}}{n^{1/4}\log n}%
\bigg\}.  \label{p3.4}
\end{equation}%
Here we took into account that $|\Re \zeta _{p}|\leq L\varepsilon <<M_{1}$.
Thus, $\Sigma _{p3}=o(n^{-1})$.

The above bounds and a bit tedious but routine calculations, based
on Proposition \ref{pro:d1} and the Euler-Mclaurin summation
formula \cite{AS} yield the r.h.s. of (\ref{teil_rho})

The same arguments can be applied also to the other endpoints of
the spectrum. Thus, Lemma \ref{pro:3} is proved.

\end{document}